\PassOptionsToPackage{sort&compress}{natbib}
\documentclass[final,3p,times]{elsarticle}
\usepackage[colorlinks,citecolor=magenta,linkcolor=blue]{hyperref}
\usepackage{graphicx}
\usepackage{natbib}
\usepackage{amsmath,amssymb}
\usepackage{txfonts,bm,listings,xcolor}
\usepackage{booktabs} 
\usepackage{algorithm}
\usepackage{algorithmic}
\usepackage{float} 

\usepackage{physics}   

\newcounter{bla}

\lstdefinelanguage{norg}{morekeywords={},
    sensitive=true,
    morecomment=[l]{\#},
    morestring=[b]",
}

\journal{Computer Physics Communications}

\begin{document}

\begin{frontmatter}

\title{\texttt{iNORG}: An open-source quantum impurity solver package based on the natural orbitals renormalization group}

\author[a]{Jia-Ming Wang}
\author[a]{Yi-Heng Tian}
\author[a]{Yin Chen}
\author[b]{Ru Zheng}
\author[a]{Rong-Qiang He\corref{author}} 
\author[a,c]{Zhong-Yi Lu\corref{author}}

\cortext[author] {Corresponding author.\\\textit{E-mail address: }rqhe@ruc.edu.cn (R.-Q. He), zlu@ruc.edu.cn (Z.-Y. Lu)}

\address[a]{School of Physics and Key Laboratory of Quantum State Construction and Manipulation (Ministry of Education), Renmin University of China, Beijing 100872, China}
\address[b]{Department of Physics, School of Physical Science and Technology, Ningbo University, Ningbo 315211, China}
\address[c]{Hefei National Laboratory, Hefei 230088, China}

\begin{abstract}
In the context of dynamical mean-field theory (DMFT) calculations for strongly correlated electron systems, quantum impurity solvers play a central computational role in treating correlated lattice models and realistic materials. Consequently, developing efficient and robust quantum impurity solvers remains a key challenge. In this paper, we present an open-source quantum impurity solver package based on the natural orbitals renormalization group (NORG) method, dubbed \texttt{iNORG}. 
This software delivers high accuracy with reduced computational cost by optimizing the bath representation using natural orbitals and incorporating advanced features such as efficient Hilbert space selection and efficient algorithms for computing Green's functions.
We first introduce the basic principle of the NORG method and then discuss the implementation details. The software framework, major features, and installation procedure for \texttt{iNORG} are explained as well. Finally, several simple examples are presented to demonstrate the usage of \texttt{iNORG}.
\end{abstract}

\begin{keyword}
Quantum impurity model \sep natural orbitals renormalization group method \sep dynamical mean-field theory 
\end{keyword}

\end{frontmatter}

\noindent {\bf PROGRAM SUMMARY}\\

\begin{small}
\noindent
{\em Program Title:} iNORG \\
{\em CPC Library link to program files:} (to be added by Technical Editor) \\
{\em Developer's repository link:} https://github.com/QTMEC-RUC/iNORG \\
{\em Code Ocean capsule:} (to be added by Technical Editor)\\
{\em Licensing provisions (please choose one):} AGPL-3.0 \\
{\em Programming language:} C++ \\
{\em Supplementary material:} \\
{\em Nature of problem (approx. 50-250 words):} Quantum many-body calculations in dynamical mean-field theory (DMFT) hinge on solving an Anderson-type impurity model embedded in a self-consistent bath. 
For realistic multi-orbital systems the bath must be discretised with tens to hundreds of orbitals, so the full Fock-space dimension rises exponentially and quickly overwhelms conventional exact-diagonalisation (ED). 
Quantum Monte-Carlo incurs a severe sign-problem at low temperature and requires ill-conditioned analytic continuation, while numerical renormalization group loses accuracy for low-temperature features with many correlated orbitals. Consequently there is a pressing need for an impurity solver that can retain ED-level accuracy for large, general-interaction impurity models yet scale far more favourably with bath size, enabling high-resolution zero-temperature DMFT studies of strongly correlated materials. \\
{\em Solution method (approx. 50-250 words):} \texttt{iNORG} addresses this challenge with the natural orbitals renormalization group (NORG). The bath is first transformed to an optimal natural-orbital basis extracted from the single-particle density matrix of the correlated ground state. 
A hierarchy of natural-orbital occupancy constraints (NOOC) compresses the space by freezing the degrees of freedom associated with nearly full or empty orbitals, thereby building a highly compressed active space. 
The reduced Hamiltonian is solved by a Lanczos algorithm, while single-particle Green's functions are obtained from continued-fraction expressions within the Krylov basis. 
Hybridization functions are discretised automatically using a Levenberg--Marquardt fit, and computation-intensive kernels call multi-threaded BLAS and LAPACK.
The overall complexity scales polynomially (\(\sim\!\mathcal{O}(N_{\mathrm{bath}}^{3})\)). \\
{\em Additional comments including restrictions and unusual features (approx. 50-250 words):} The current version of \texttt{iNORG} is restricted to zero-temperature, single-site impurity models and presently supports only density-density Coulomb interactions; extensions for finite-temperature calculations, cluster DMFT, and two-particle response functions are planned for future releases. Among its notable features is an adaptive natural orbital occupancy constraint (NOOC) scheme that dynamically adjusts the active Hilbert space to optimize computational efficiency. The implementation introduces a `NORG-Table' data structure that caches sparse matrix information, enabling rapid Hamiltonian reconstruction across self-consistent iterations. Furthermore, the automated bath discretization routine provides advanced user control, allowing for non-uniform frequency grids to be specified for the hybridization function fitting, which is particularly useful for resolving sharp spectral features.
\end{small}

\section{Introduction\label{sec:intro}}
In recent years, strongly correlated electron systems, such as high-temperature superconductors and quantum magnets, have garnered significant attention due to their rich physical phenomena and potential applications in advanced materials. A central challenge in this field is accurately describing the behavior of electrons in the presence of strong interactions, which generally requires solving complex quantum many-body problems. One effective approach to tackle these systems is dynamical mean-field theory (DMFT), which maps a lattice problem onto an effective impurity model embedded in a self-consistently determined bath \cite{metzner1989correlated,georges1996dynamical}.

A central component of DMFT is the impurity solver, which computes the self-energy of the impurity model. The Anderson impurity model (AIM) \cite{anderson1961localized}, which describes localized electrons with Coulomb interactions in a conduction bath, is fundamental to studying strongly correlated systems. Generally, it underpins phenomena like the Kondo effect \cite{kondo1964resistance}, dilute magnetic impurities \cite{hewson1997book}, and impurity quantum phase transitions \cite{Vojta2006tf}. In DMFT \cite{metzner1989correlated,georges1996dynamical}, the AIM plays a central role, as lattice models are mapped to self-consistent impurity problems, with multi-orbital systems posing significant computational challenges due to complex inter-orbital hybridizations \cite{maier2005quantum,surer2012multiorbital}.

Analytical solutions for the AIM are typically infeasible, necessitating numerical methods. Various impurity solvers have been developed, each with its strengths and limitations. Traditional approaches include exact diagonalization (ED) \cite{noack2005diagonalization} and Lanczos \cite{lin1993exact}, which are accurate for small bath sizes but scale exponentially with Hilbert space size; quantum Monte Carlo (QMC) \cite{gull2011continuous}, suited for finite temperatures but limited at low temperatures or real frequencies; the numerical renormalization group (NRG) \cite{wilson1975renormalization}, effective for low energies but less so for high energies or multi-orbital systems; and hierarchical equations of motion (HEOM) \cite{li2012hierarchical}, which is efficient under specific conditions but costly with increasing bath complexity.

To address these challenges, the natural orbitals renormalization group (NORG) method has been proposed as an efficient and accurate impurity solver \cite{he2014prb,wang2025abinitio}. NORG leverages the concept of natural orbitals (NOs) \cite{lowdin1955quantum,bender1966natural} to find a compact ground-state representation for the AIM, exploiting correlation sparsity. This enables efficient approximations with few Slater determinants, reducing computational complexity. In this article, we introduce \texttt{iNORG}, an open-source implementation of the NORG method designed to solve quantum impurity models effectively, particularly for multi-orbital systems where other methods struggle.

\section{Model and methods\label{sec:basic}}
This section first introduces the Anderson impurity model solved by \texttt{iNORG} and explains the motivation for optimizing the bath-orbital representation using natural orbitals. Subsequently, we present the definition of natural orbitals and orbital transformations, and detail the self-consistent procedure of the iterative natural orbitals renormalization group (NORG) method and subspace screening strategies. Finally, we describe the Fock-space encoding of many-body wave functions and the Lanczos ground-state solver, which provide the algorithmic basis for the implementation and optimizations discussed below.

\subsection{Anderson impurity model\label{subsec:impmodel}}
Within the dynamical mean-field theory (DMFT), solving the Anderson impurity model (AIM)~\cite{anderson1961localized} constitutes the computational core. The AIM describes a localized interacting subsystem (the impurity) coupled to a non-interacting bath, and it is widely studied for strongly correlated electron phenomena, such as the Kondo effect and Mott transitions. Typically, the bath represents conduction electrons. The general form of the Hamiltonian can be expressed as:
\begin{equation}
H_{\text{imp}} = H_{\text{loc}} + H_{\text{bath}} + H_{\text{hyb}}\,,
\end{equation}
where
\begin{equation}
  \begin{aligned}
    H_{\text{loc}} &= \sum_{\alpha\beta} E_{\alpha\beta} d_{\alpha}^{\dagger} d_{\beta} + \sum_{\alpha\beta\gamma\delta} U_{\alpha\beta\gamma\delta} 
        d^{\dagger}_{\alpha} d^{\dagger}_{\beta} d_{\gamma} d_{\delta}\,, \\
    H_{\text{hyb}} &= \sum_{\textbf{k}\alpha\beta} V^{\alpha\beta}_{\textbf{k}} d_{\alpha}^{\dagger} c_{\beta \textbf{k}} + \text{h.c.}\,, \\
    H_{\text{bath}} &= \sum_{\alpha\textbf{k}} \epsilon_{\alpha\textbf{k}} c_{\alpha\textbf{k}}^{\dagger} c_{\alpha\textbf{k}}\,.
  \end{aligned}
  \label{eq1:aim}
\end{equation}
Here, $d_{\alpha}^{\dagger}$ and $c_{\alpha\textbf{k}}^{\dagger}$ are the electron creation operators for the impurity and bath, respectively, with $\alpha$ and $\beta$ being spin-orbital indices. $H_{\text{loc}}$ accounts for both the single-particle energies and Coulomb interactions on the impurity orbitals. $H_{\text{hyb}}$ characterizes the hybridization between the impurity and bath orbitals, while $H_{\text{bath}}$ corresponds to the non-interacting conduction bath. 
Solving the AIM requires balancing computational speed and accuracy. Three main solvers span this spectrum: the Hubbard-I approximation, the exact diagonalization (ED), and the quantum Monte Carlo (QMC). Although the model itself is compact, reliable Green's functions must be obtained from computationally intensive numerical methods. The Hubbard-I runs quickly but misses strong-correlation physics \cite{hubbard1963electron}. The ED yields exact results for small clusters but its cost grows exponentially with system size \cite{caffarel1994exact}. The QMC can handle larger systems, yet it suffers from the sign problem and the need to analytically continue data from imaginary to real frequencies \cite{gull2011continuous}. Modern DMFT studies demand many orbitals, very low temperatures, and high real-frequency resolution. No single solver can yet meet all these requirements simultaneously.

This computational bottleneck has motivated the development of the natural orbitals renormalization group (NORG) approach~\cite{he2014prb,wang2025abinitio}, which circumvents many traditional limitations through a fundamentally different strategy: 
the iterative optimization of the orbitals themselves for a fixed many-body wave function that complies with natural orbital constraints.
By fully utilizing the inherent sparsity in the AIM and focusing computational resources on the most quantum-mechanically active degrees of freedom while effectively treating less relevant contributions, the NORG achieves a remarkable balance between accuracy and efficiency, which opens new possibilities for tackling previously intractable many-body problems. In the last few years, the NORG method has been successfully used as an impurity solver in DMFT for many projects, such as the investigation of low-energy inter-band Kondo bound states in orbital-selective Mott phases~\cite{wang2025low}, the DFT+DMFT analysis of correlated electronic structures in layered nickelates~\cite{ouyang2025dft}, the study of non-Fermi-liquid regimes and antiferromagnetic correlations in bilayer Hubbard models~\cite{chen2024nonfermi}, and the elucidation of two-orbital s$_\pm$-wave superconductivity in high-pressure La$_3$Ni$_2$O$_7$~\cite{tian2024correlation}.

\subsection{Natural orbitals renormalization group method\label{subsec:norg}}
\subsubsection{Natural orbitals\label{subsec:no}}
Natural orbitals (NOs) were first proposed by Löwdin in chemistry~\cite{lowdin1955quantum,lowdin1956natural}. We demonstrated that natural orbitals provide an optimal single-particle basis for describing the many-body quantum state $|\Psi\rangle$ of the AIM \cite{he2014prb}. They are obtained by diagonalizing the single-particle density matrix (SPDM), whose elements are
\begin{equation}
D_{ij} = \langle \Psi | \hat{c}_i^\dagger \hat{c}_j | \Psi \rangle,
\end{equation}
where $\hat{c}_i^\dagger$ and $\hat{c}_j$ are creation and annihilation operators in an arbitrary initial single-particle basis $\{|i\rangle\}$, respectively. The eigenvectors of the SPDM define the NOs, denoted as $\{|k\rangle\}$, and the corresponding eigenvalues are the natural orbital occupation numbers $n_k$. For a spin-orbital basis, each spin-orbital can be occupied by at most one fermion, so $0 \le n_k \le 1$. When spin is not resolved, each orbital groups the spin-up and spin-down components, and its occupation satisfies $0 \le n_k \le 2$.

In the framework of the NORG method~\cite{he2014prb,wang2025abinitio}, the NO basis is iteratively refined to optimize the representation of a many-body wave function, enabling efficient solutions of complex quantum systems. To formulate the NORG method, we first transform the initial single-particle basis into a general orbital basis $\mathcal{S}$. In this basis, the single-particle orbitals are orthonormal and denoted as $|g\rangle = d_g^\dagger |\text{vac}\rangle$ with $g = 1, 2, \dots, L$, where $|\text{vac}\rangle$ is the vacuum state and $d_g^\dagger$ and $d_g$ are the creation and annihilation operators for the $g$-th orbital. The transformation to this basis can be written as $|g\rangle = \sum_{i=1}^L U_{gi}^\dagger |i\rangle$, where $U_{gi}$ are the elements of a unitary matrix $U$. Then, the corresponding operator transformations are expressed as:
\begin{equation}\label{eq2:transformation}
  d_g^\dagger = \sum_{i=1}^L U_{gi}^\dagger c_i^\dagger, \quad d_g = \sum_{i=1}^L U_{ig} c_i,
\end{equation}
or, in the matrix form:
\[
    \begin{pmatrix}
    d_1^\dagger \\
    d_2^\dagger \\
    \vdots \\
    d_L^\dagger
    \end{pmatrix}
    = U^\dagger
    \begin{pmatrix}
    c_1^\dagger \\
    c_2^\dagger \\
    \vdots \\
    c_L^\dagger
    \end{pmatrix},
    \quad
    \begin{pmatrix}
    d_1, & d_2, & \cdots, & d_L
    \end{pmatrix}
    =
    \begin{pmatrix}
    c_1, & c_2, & \cdots, & c_L
    \end{pmatrix}
    U.
\]
Since $U$ is unitary, the new operators satisfy the fermionic anticommutation relations: $\{ d_g, d_{g'}^\dagger \} = \delta_{gg'}$, $\{ d_g, d_{g'} \} = 0$, $\{ d_g^\dagger, d_{g'}^\dagger \} = 0$. Correspondingly, the inverse transformation can be obtained by:
\begin{equation}
  c_i^\dagger = \sum_{g=1}^L U_{ig} d_g^\dagger, \quad c_i = \sum_{g=1}^L U_{gi}^\dagger d_g.
\end{equation}

In the representation $\mathcal{S}$ (which NORG aims to optimize), the wave function $\ket{\Psi}$ can be expanded by considering the occupation of a specific orbital $g$:
\begin{equation}\label{eq2:ketPsi}
  \ket{\Psi} = \sum_i h_i^\circ \ket{\phi_i^{\circ,g}} + \sum_j h_j^\bullet \ket{\phi_j^{\bullet,g}},
\end{equation}
where $\ket{\phi_i^{\circ,g}}$ are $N$-electron Slater determinants not containing the $g$-th orbital, and $\ket{\phi_j^{\bullet,g}}$ are those containing the $g$-th orbital. Thus Eq.~\eqref{eq2:ketPsi} partitions the wave function into the component in which orbital $g$ is empty and the component in which it is occupied. These Slater determinants are orthonormal:
\[
  \langle \phi_i^m | \phi_j^n \rangle = \delta_{ij} \delta_{mn}, \quad m, n \in \{\circ, \bullet\},
\]
and the coefficients satisfy the normalisation condition:
\[
  \sum_i |h_i^\circ|^2 + \sum_j |h_j^\bullet|^2 = 1.
\]
The occupation number of orbital $g$ is
\begin{equation}
  n_g^d \equiv \langle \Psi | d_g^\dagger d_g | \Psi \rangle.
\end{equation}
Using Eq.~\eqref{eq2:ketPsi}, we obtain
\begin{equation}
  n_g^d = \sum_j |h_j^\bullet|^2 = 1 - \sum_i |h_i^\circ|^2,
\end{equation}
which shows that $n_g^d$ directly measures the occupation of orbital $g$ in $\ket{\Psi}$. The NORG method iteratively optimizes $\mathcal{S}$ (by adjusting $U$) so that most transformed orbitals have occupations close to either 0 or 1. This separates active degrees of freedom, which carry large quantum fluctuations, from inactive ones, which can be treated approximately to balance accuracy and computational efficiency.

In the NORG framework, orbitals in the optimized basis $\mathcal{S}$ are classified by their occupation numbers $n_g^d$:
\begin{itemize}
  \item \textbf{Inactive orbitals}: $n_g^d \approx 0$ (nearly empty) or $n_g^d \approx 1$ (nearly full). In the spin-orbital basis used here, each orbital is a single fermionic mode, so $0 \le n_g^d \le 1$. Because nearly empty or nearly full orbitals contribute weakly to the fluctuating part of the many-body wave function, they can be ``frozen'' to reduce the Hilbert space dimension, with an associated error of approximately $n_g^d$ or $(1 - n_g^d)$.
  \item \textbf{Active orbitals}: Orbitals with $n_g^d$ significantly different from 0 and 1, indicating substantial quantum fluctuations. These require explicit treatment for an accurate system description.
\end{itemize}
This classification thus partitions the NO basis into active and inactive subsets. The NORG method leverages this by iteratively optimizing the basis $\mathcal{S}$ to distinguish between active and inactive NOs, enabling a compressed representation of the many-body wave function $\ket{\Psi}$. The key advantage is that $\ket{\Psi}$ converges more rapidly in this basis, allowing computational effort to focus on active NOs while approximating inactive ones. Hence, this reduction in Hilbert space complexity is the cornerstone of NORG's efficiency.

\subsubsection{Renormalization strategy\label{subsec:rg}}
NORG is a non-perturbative numerical technique tailored for quantum impurity models that effectively exploits the NO basis~\cite{he2014prb,wang2025abinitio}. It is especially effective for systems with sparse interactions, as is typical in the AIMs from DMFT, in which interactions are confined to impurity sites.

The core NORG workflow is an iterative self-consistent procedure:
\begin{enumerate}
    \item \textbf{Initialize}: Start from an initial set of single-particle orbitals (e.g., the original bath sites).
    \item \textbf{Construct subspace}: Apply different constraints to different orbitals to construct an effective Hilbert subspace.
    \item \textbf{Build effective Hamiltonian}: Construct the Hamiltonian matrix $\hat{H}_{\text{eff}}$ within the chosen active subspace.
    \item \textbf{Solve ground state}: Solve for the approximate ground state $|\Psi_{\text{sub}}\rangle$ of $\hat{H}_{\text{eff}}$, typically using an iterative method such as Lanczos.
    \item \textbf{Compute SPDM}: Compute the SPDM $D_{\text{sub}}$ using the obtained $|\Psi_{\text{sub}}\rangle$.
    \item \textbf{Update NOs}: Diagonalize $D_{\text{sub}}$ to obtain a new set of NOs and their occupations $n_k$.
    \item \textbf{Iterate}: Use the updated NOs to refine the selection of the active subspace and repeat steps 3--7 until convergence criteria (e.g., ground state energy, occupations) are met.
\end{enumerate}
By iteratively refining the NO basis and dynamically adjusting the active space to focus on the most relevant quantum fluctuations (active NOs) while freezing the less important ones (inactive NOs), NORG achieves high accuracy with significantly reduced computational cost compared to the ED of the full problem. The computational complexity scales polynomially with the number of bath sites $N_{\text{bath}}$ (typically $O(N_{\text{bath}}^3)$ or $O(N_{\text{bath}}^4)$), making it feasible to handle systems with hundreds of bath orbitals.

In the NORG computational process, the model is transformed from a lattice basis (the real-space representation) to the natural orbital basis (the optimized basis composed of the eigenstates of the single-particle density matrix). The orbital transformation aims to improve computational efficiency through the optimized NO basis. Two transformation strategies can be adopted:
\begin{enumerate}
    \item \textbf{Transforming all orbitals simultaneously:} The entire single-particle density matrix (SPDM) is diagonalized to obtain a unitary transformation $U$ acting on all orbitals. All sites in the original model (including interacting impurity orbitals and non-interacting bath orbitals) are mapped to the NO space. However, interaction terms (such as the local Hubbard $U$) are then distributed across multiple NOs, increasing the number of non-zero elements in the Hamiltonian matrix and making it denser. The computational complexity grows as $O(N^4)$ with the total number of orbitals $N$.
    \item \textbf{Transforming only non-interacting orbitals:} Only the block of the SPDM corresponding to the bath orbitals (non-interacting sites) is diagonalized, while the impurity orbitals (interacting sites) are left unchanged. This strategy avoids distributing the interaction terms, keeping the Hamiltonian matrix sparse. Although this transformation may introduce weak off-diagonal couplings among the new bath NOs, the number of non-zero Hamiltonian elements increases only slightly. The computational complexity grows as $O(N_{\text{bath}}^3)$ with the number of bath orbitals $N_{\text{bath}}$, significantly improving efficiency. The full-orbital transformation is more general and can be useful for small systems or diagnostic analyses of the complete SPDM, but for AIMs with local impurity interactions the bath-only transformation preserves the local interaction structure and is therefore the default strategy in \texttt{iNORG}.
\end{enumerate}

\subsection{Representation of many-body wave functions\label{subsec:fockstate}}

In the second quantization framework, many-electron wave functions are typically represented using Fock states. For a system with \(M\) spatial orbitals, each comprises two spin degrees of freedom (up and down), resulting in a total of \(2M\) spin‑orbitals. By assigning an occupation number \(n_i \in \{0,1\}\) to each orbital, the complete many-body Hilbert space can be constructed as:
\[
  \mathcal{H} = \mathcal{H}_0 \oplus \mathcal{H}_1 \oplus \cdots \oplus \mathcal{H}_{2M},
\]
where \(\mathcal{H}_N\) represents the subspace with electron occupation number \(N\), and the total dimension is \(2^{2M}\). Since the Hilbert space dimension grows exponentially with the number of orbitals \(2M\), practical calculations usually focus only on the subspace \(\mathcal{H}_N\) with a specific occupation number \(N\).

Within this framework, creation and annihilation operators \(\hat{c}_i^\dagger\) and \(\hat{c}_i\) flip the occupation of orbital \(i\) (\(i = 0, 1, \dots, 2M-1\)) between 0 and 1.
This "0-1" binary nature reflects the Pauli exclusion principle, i.e., each orbital can be occupied by at most one fermion.

In the \texttt{iNORG} code, a single-particle basis \(\{\alpha\}\) is first fixed, and based on this, the Fock basis \(\{|I\rangle\}\) is constructed. The orbital numbering follows the convention: spin-up (\(\uparrow\)) orbitals first, followed by spin-down (\(\downarrow\)) orbitals. Specifically, orbitals are numbered from 0 to \(2M-1\), where:
\[
  0, 1, \dots, M-1 \quad (\uparrow \text{ states}); \qquad M, M+1, \dots, 2M-1 \quad (\downarrow \text{ states}).
\]

Fock states are encoded using a \(2M\)-bit binary number, where "1" indicates the orbital is occupied, and "0" indicates it is unoccupied. For example, consider a system with \(M=3\), having 6 orbitals numbered 0 (\(\uparrow\)), 1 (\(\uparrow\)), 2 (\(\uparrow\)), 3 (\(\downarrow\)), 4 (\(\downarrow\)), 5 (\(\downarrow\)). The binary encoding convention has the least significant bit (LSB) corresponding to orbital 0, increasing to orbital 5. A specific Fock state binary string is:
\[
  \overbrace{\color{red}{001}}^{\text{0--2}}
  \overbrace{\color{blue}{100}}^{\text{3--5}},
\]
indicating that orbital 2 (\(\uparrow\)) and orbital 3 (\(\downarrow\)) are occupied. Its corresponding decimal value is:
\begin{equation}
\begin{aligned}
  &0 \times 2^0 + 0 \times 2^1 + 1 \times 2^2 + 1 \times 2^3 + 0 \times 2^4 + 0 \times 2^5 \\
  &= 0 + 0 + 4 + 8 + 0 + 0 = 12.
\end{aligned}
\end{equation}
This binary encoding facilitates computer storage and manipulation, providing an efficient representation for numerical solutions of many-body problems.

\subsection{Lanczos algorithm\label{subsec:lanc_alg}}

The Lanczos algorithm has been thoroughly described in multiple review articles \cite{weisse2006rmp, jaklic1994prb}. This section briefly outlines the basic idea of the algorithm and provides its pseudocode. The Lanczos algorithm is suitable for computing the extremal eigenvalues and eigenvectors of large, sparse matrices using only repeated Hamiltonian-vector products, and can also be used to compute spectral functions (such as single-particle Green's functions), which we will introduce in the following sections. In this section, we focus exclusively on ground-state calculation.

The core idea of the Lanczos method is to avoid direct diagonalization of the full Hamiltonian \(H\). Instead, starting from a random initial state vector \(\ket{\varphi_0}\), it iteratively constructs an orthonormal basis for a much smaller subspace, known as the Krylov subspace:
\begin{equation}
\mathcal{K}_m(H, \ket{\varphi_0}) = \text{span}\{\ket{\varphi_0},\, H\ket{\varphi_0},\, H^2\ket{\varphi_0},\, \ldots,\, H^{m-1}\ket{\varphi_0}\}.
\end{equation}
In this basis, the projection of the original Hamiltonian \(H\) becomes a small, tridiagonal matrix, \(T_m\). The extremal eigenvalues of \(T_m\) usually converge rapidly to those of the full \(H\). By diagonalizing this small tridiagonal matrix, one can obtain a highly accurate approximation of the ground-state energy and wave function of the original system with a modest number of iterations (typically \(m \approx 200\)) \cite{wu1999jcp, sundar2000cma, wu2000siam, kokiopoulou2004anm}.

The orthonormal basis vectors \(\{\ket{\varphi_i}\}\) that tridiagonalize the Hamiltonian are generated via a three-term recurrence relation. Starting with a normalized random vector \(\ket{\varphi_0}\), the sequence is generated as follows:
\begin{equation}
  \begin{split}
  \alpha_i &= \langle \varphi_i | H | \varphi_i \rangle \\
  \beta_{i+1} \ket{\varphi_{i+1}} &= (H - \alpha_i) \ket{\varphi_i} - \beta_i \ket{\varphi_{i-1}}.
  \end{split}
  \label{eq:lancKryl}
\end{equation}
where \(\beta_0=0\) and \(\ket{\varphi_{-1}}=0\). In each step, the new vector \(\ket{\varphi_{i+1}}\) is orthogonalized against the previous two vectors and then normalized. The coefficients \(\{\alpha_i\}\) and \(\{\beta_i\}\) form the diagonal and off-diagonal elements, respectively, of the tridiagonal matrix \(T_m\).

\section{Implementations and optimizations\label{sec:opt}}
\subsection{Overall workflow of \texttt{iNORG}}

\begin{figure}[H]
  \centering
  \includegraphics[width=\textwidth]{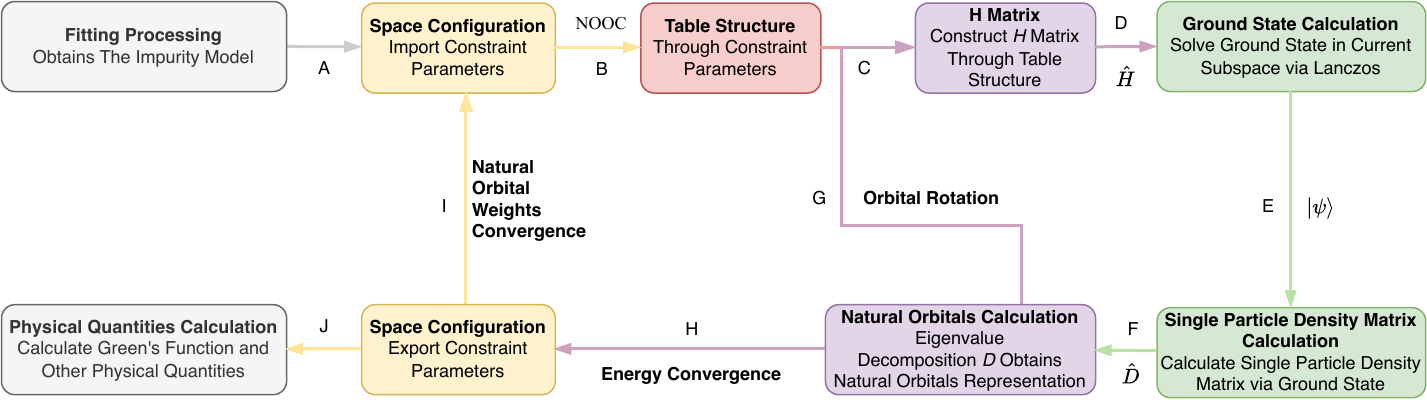}
  \caption{Flowchart of the natural orbitals renormalization group (NORG) method as implemented in the \texttt{iNORG} package. The process starts from an Anderson impurity model defined by DMFT (A). It then enters a self-consistent cycle (B-I) to optimize the natural orbitals by repeatedly constructing and solving the Hamiltonian within a selected Hilbert subspace. After convergence, physical observables are calculated (J).}
  \label{fig:iNORG-flowchart}
\end{figure}

This section describes the implementation details and optimization techniques of \texttt{iNORG}, an open-source C++ software package \cite{QTMEC-RUC-iNORG} specifically designed for impurity solving within the DMFT framework.
It employs the natural orbitals renormalization group (NORG) method~\cite{he2014prb} to efficiently handle strongly correlated electron behavior in quantum impurity models. Figure \ref{fig:iNORG-flowchart} illustrates the main computational workflow of \texttt{iNORG}, which we will explore in detail in the following sections.

\begin{enumerate}
  \item[Step A:] Obtain the local hybridization function \(\varGamma_{\rm loc}(z)\) via DMFT, which characterizes the interaction between the impurity and the environmental bath. The preprocessing module discretizes the continuous \(\varGamma_{\rm loc}(z)\) into a finite number of bath sites. In the Anderson Impurity Model (\( H_{\text{imp}} = H_{\text{loc}} + H_{\text{bath}} + H_{\text{hyb}} \), see Eq.~\eqref{eq1:aim}), this determines \(H_{\text{hyb}}\) and \(H_{\text{bath}}\), ensuring the impurity's hybridization function \(\varGamma_{\rm imp}(z) \approx \varGamma_{\rm loc}(z)\). Interaction parameters (e.g., Coulomb interaction strength) are input to define \(H_{\text{loc}}\), thus specifying the impurity model.
  \item[Step B:] Screen the Hilbert subspace using the natural orbital occupancy constraint (NOOC) algorithm, constructing the NORG-Table structure required for computation.
  \item[Step C:] Utilize the NORG-Table structure to store the mapping between many-body bases in the Hilbert subspace and operator matrix elements in the single-particle basis, eliminating the need to construct various many-body operator matrices during subsequent iterations.
  \item[Step D:] Apply the Lanczos algorithm to the Hamiltonian \(\hat{H}\) to obtain the ground-state wave function \(\ket{\psi}\) and ground-state energy \(E_0\) within the selected subspace.
  \item[Step E:] Compute the single-particle density matrix \(D_{\alpha \beta} = \left\langle \psi \right| c_\alpha^{\dagger} c_\beta \left| \psi \right\rangle\) based on \(\ket{\psi}\).
  \item[Step F:] Perform eigenvalue decomposition on \(D\) to obtain natural orbitals \(\{\phi_i\}\) and their occupancies \(\{n_i\}\).
  \item[Step G:] Check whether the ground-state energy difference between two consecutive NORG iterations satisfies the convergence criterion. If not, return to Step C to optimize the subspace; otherwise, proceed to Step H.
  \item[Step H:] Export the occupancy numbers in the natural orbital representation and record the natural orbital transformation matrix.
  \item[Step I:] Calculate the norm difference of natural orbital occupancies between consecutive iterations, \(d(n_{\rm i,last}, n_{\rm i,new}) = \sum_{i=1}^n |n_{\rm i,last} - n_{\rm i,new}|\). If \(d \leq \Delta_d\), proceed to Step J; otherwise, return to Step B to adjust the Hilbert subspace.
  \item[Step J:] After natural orbital convergence, compute the Green's function and other physical quantities.
\end{enumerate}

To realize this workflow, \texttt{iNORG} adopts the following technical strategies during development to ensure flexibility, stability, and extensibility:
\begin{enumerate}
  \item Modular and Independent Design: All modules are written in C++17 using object-oriented programming, allowing independent operation while reducing coupling for easy maintenance and expansion.
  \item Development Environment: Targeted for Linux platforms, using Makefiles for compilation management, integrated with MPI parallelism and Intel oneAPI's BLAS and LAPACK libraries.
  \item Model Support: Defaults to DMFT calculations for multi-orbital Bethe lattices, with the ability to support other lattice models through modifications.
  \item Software Integration: Compatible with DFT+DMFT software (e.g., ZEN, eDMFT), enabling parameter passing via configuration files for seamless data exchange and collaborative computation.
\end{enumerate}

\begin{figure}[H]
  \centering
  \includegraphics[width=10cm]{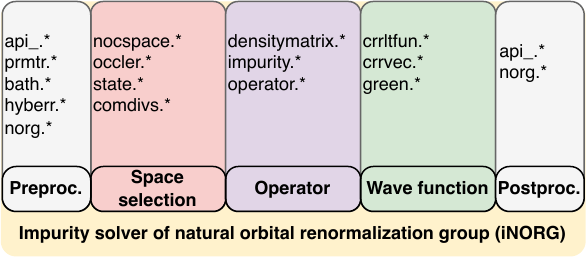}
  \caption[Organizational structure of the \texttt{iNORG} program]{Organizational structure of the \texttt{iNORG} program. The first row shows the module abbreviations, and the second row indicates the corresponding module functions.}
  \label{fig:3-codestructure}
\end{figure}

The implementation of these technical strategies, especially the principle of modular design, results in the software architecture illustrated in Fig.~\ref{fig:3-codestructure}. This diagram outlines the core modules of \texttt{iNORG} and uses color-coding to visually link each module to its corresponding steps in the workflow flowchart (as shown in Fig.~\ref{fig:iNORG-flowchart}).

The functions of these modules are summarized below. Owing to space constraints, the following subsections present only the essential technical details required to reproduce the program, omitting repetitive descriptions of similar implementation steps.
\begin{itemize}
  \item \textbf{Preprocessing and Postprocessing}: Implements hybridization function fitting (Step A) and physical quantity computation (Step J), detailed in Section \ref{subsec:hybfit}.
  \item \textbf{Space selection}: Includes NOOC (Step B), operator relation storage (Step C), and convergence judgment (Step I), detailed in Sections \ref{subsec:NOOC}, \ref{subsec:fockstate}, and \ref{subsec:wavefunctioncode}.
  \item \textbf{Operator}: Rapid construction of many-body operators (Step D) and orbital transformations (Step F), detailed in Sections \ref{subsec:operator}.
  \item \textbf{Wave function}: Ground-state solution via the Lanczos algorithm (Step D), wave function analysis (Step E), and physical quantity computation (Steps H, J), detailed in Sections \ref{subsec:lanc_gs}, \ref{subsec:continuefraction}, and \ref{subsec:krylovsubspace}.
\end{itemize}

\subsection{Hybridization function fitting algorithm\label{subsec:hybfit}}
In this software package, we have implemented an automated hybridization function processing capability, employing the Levenberg--Marquardt fitting algorithm to discretize the impurity model within the DMFT framework. In the context of strongly correlated electron systems under DMFT, the continuous local hybridization function \(\varGamma_{\rm loc}(z)\) must be discretized into a finite number of bath sites to construct an effective model coupled to the impurity orbitals. This process can be viewed as a nonlinear least-squares problem, aiming to minimize the error between the fitted hybridization function \(\varGamma_{\rm imp}(z)\) and the original \(\varGamma_{\rm loc}(z)\), where \(\varGamma_{\rm loc}(z) = z - \mathcal{G}_l^{-1}(z)\) is defined by the local Green's function \(\mathcal{G}_l(z)\).

\subsubsection{Fitting details}
The Levenberg--Marquardt algorithm combines the strengths of the Gauss--Newton method and gradient descent, using a damping factor \(\lambda\) to balance numerical stability and convergence speed. In each iteration, the parameter update equation is:
\begin{equation}
  \boldsymbol{\theta}_{\mathrm{new}} = \boldsymbol{\theta}_{\mathrm{old}} - \bigl(J^\mathrm{T} J + \lambda I \bigr)^{-1} J^\mathrm{T} \mathbf{r},
\end{equation}
where \(\mathbf{r}\) is the residual vector, and \(J\) is the Jacobian matrix. The damping factor \(\lambda\) adjusts dynamically based on fitting performance: when the residual \(\chi^2\) decreases, \(\lambda\) is reduced to accelerate convergence; when \(\chi^2\) increases, \(\lambda\) is increased to ensure stability. The algorithm exhibits near-quadratic convergence near the optimal solution, effectively reducing iteration counts and enhancing discretization efficiency and accuracy~\cite{wang2023cg}.

The choice of initial parameters is crucial for convergence. Leveraging physical prior knowledge (e.g., spectral symmetry or hybridization strength estimates) to set the initial bath energy levels \(\epsilon_k\) and hybridization amplitudes \(V_k\) reduces the risk of converging to local minima. In numerical implementation, we monitor the condition number of the Jacobian matrix to ensure stability and employ stopping criteria (e.g., \(\chi^2\) change below a threshold or maximum iteration limit) to guarantee robust and efficient fitting.

Specifically, the discretized hybridization function is expressed as:
\begin{equation}
  \varGamma_{\rm imp}(z) = \sum_{k=1}^{N_b} \frac{|V_k|^2}{z - \epsilon_k},
\end{equation}
where \(N_b\) is the number of bath sites. On discrete complex frequency points \(\{z_n\}_{n=1}^{N_z}\)—e.g., Matsubara frequencies \(z_n = i\omega_n\)—the target residual function is:
\begin{align}
  \chi^{2} &= \sum_{n} w_{n} \left| \varGamma_{\mathrm{loc}}(z_{n}) - V (\epsilon_{\text{bath}} - z_{n})^{-1} V^{\dagger} \right|^{2} \notag \\
  &\quad + w_{N+1} \left| V V^{\dagger} - (K_{0,\text{loc}} - (H_{0,\text{loc}})^{2}) \right|^{2} \notag \\
  &\quad + w_{N+2} \left( \sum_{k} \epsilon_{\text{bath},k}^{4} \right)^{2},
\end{align}
where the first term measures the fitting error between \(\varGamma_{\mathrm{loc}}(z_n)\) and the discretized hybridization function, which is formed from the impurity-bath coupling matrix \(V\) and the bath energy levels \(\epsilon_{\text{bath}}\) as \(V (\epsilon_{\text{bath}} - z_n I)^{-1} V^{\dagger}\). The second term introduces bath sum rules~\cite{koch2008prb}. At high frequency the discretized hybridization falls off as \(1/z\), and the coefficient of this leading term is governed by \(V V^{\dagger}\). Penalizing the deviation of \(V V^{\dagger}\) from the local non-interacting variance \(K_{0,\text{loc}} - (H_{0,\text{loc}})^{2}\) (the second central moment of \(H_0\) over the Brillouin zone) therefore constrains the fitted bath to reproduce the correct high-frequency (\(1/z\)) behavior of the exact hybridization. The moments $H_{0, \text{loc}}$ and $K_{0, \text{loc}}$ are defined by integrating the non-interacting Hamiltonian $H_0$ and its square over the Brillouin zone: $H_{0, \text { loc }} \equiv \int_{\mathrm{BZ}} \frac{\mathrm{d} k^d}{(2 \pi)^d} H_0$ and $K_{0, \text { loc }} \equiv \int_{\mathrm{BZ}} \frac{\mathrm{d} k^d}{(2 \pi)^d} (H_0)^2$. The third term acts as a normalization constraint, limiting the fourth power of bath energy levels \(\epsilon_{\text{bath},k}\) to prevent parameter divergence. While a simple function is used here as an example, normalization constraints can also take more complex forms.

Ultimately, the impurity Hamiltonian is constructed as:
\begin{equation}
  H_{\rm imp} = H_{\rm loc} + H_{\rm bath} + H_{\rm hyb},
\end{equation}
where:
\begin{equation}
  H_{\rm bath} = \sum_{\alpha\textbf{k}} \epsilon_{\alpha\textbf{k}} c_{\alpha\textbf{k}}^{\dagger} c_{\alpha\textbf{k}},
\end{equation}
\begin{equation}
  H_{\rm hyb} = \sum_{\textbf{k}\alpha\beta} V^{\alpha\beta}_{\textbf{k}} d_{\alpha}^{\dagger} c_{\beta\textbf{k}} + \text{h.c.},
\end{equation}
and \(H_{\rm loc}\) is the local part of the original lattice model, written exactly as in Eq.~\eqref{eq1:aim}. The operators follow the same convention as Eq.~\eqref{eq1:aim}: \(d_{\alpha}^{\dagger}\) creates an impurity electron and \(c_{\alpha\textbf{k}}^{\dagger}\) creates a bath electron, with \(\alpha,\beta\) spin-orbital indices. The only difference is that the continuous conduction bath of Eq.~\eqref{eq1:aim} is replaced by a finite set of discrete bath sites labeled by \(\textbf{k}\), whose energies \(\epsilon_{\alpha\textbf{k}}\) and hybridization amplitudes \(V^{\alpha\beta}_{\textbf{k}}\) are determined by the discretization fit. This model accurately reproduces local dynamical properties within DMFT iterations.

\subsubsection{Fitting performance}
\begin{figure}[H]
  \centering
  \includegraphics[width=8cm]{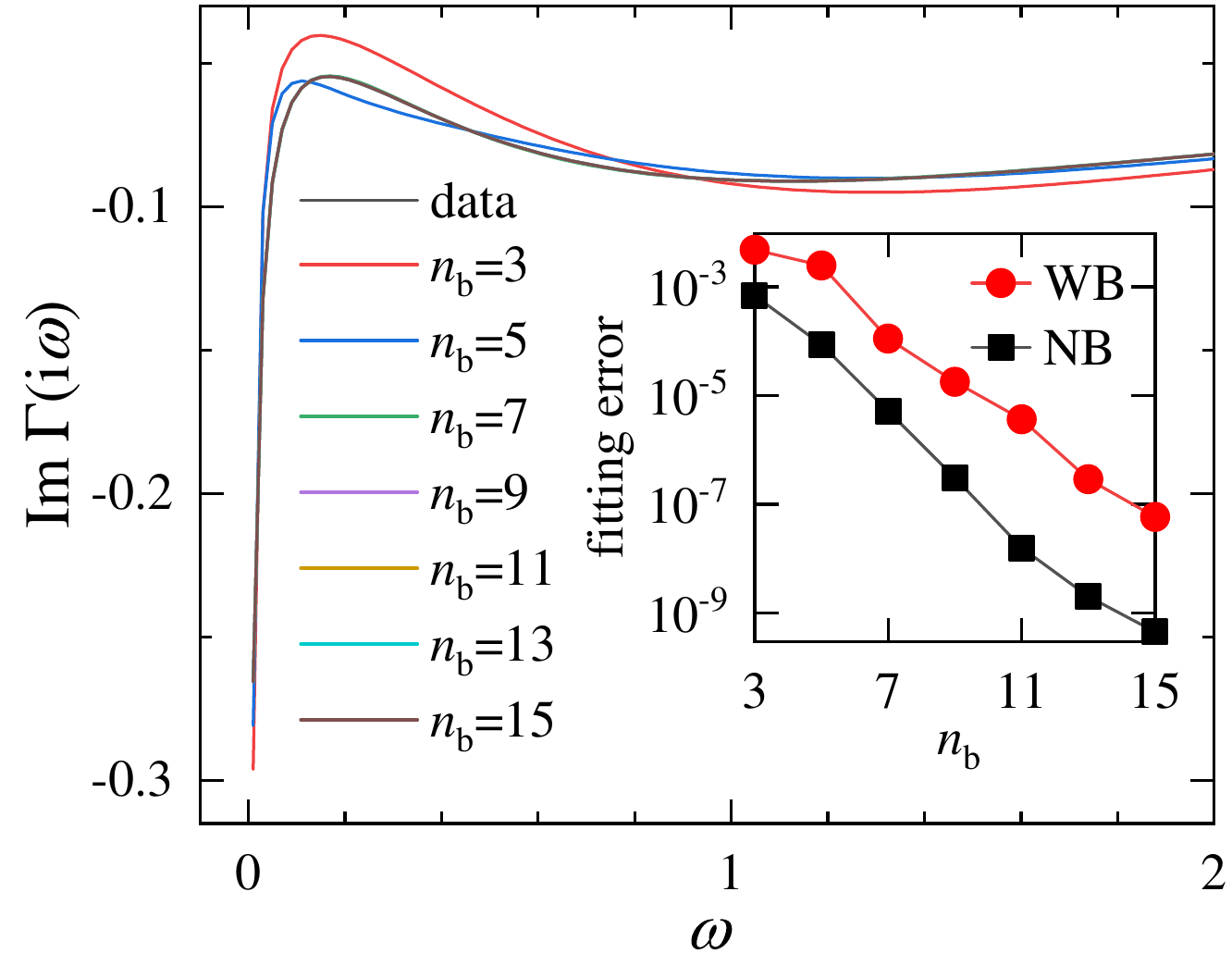}
  \caption[Impurity Model Fitting Performance Example]{Impurity Model Fitting Performance Example: Displays the imaginary part of the wide-band (WB) hybridization function \(\varGamma_{\rm imp}(i \omega)\), with parameters \(U = 3\), \(\Delta = 0.3\), \(t_2 = 0.5 t_1\). The inset shows fitting errors for wide-band (WB) and narrow-band (NB) cases under varying \(n_{\rm b}\) (number of bath sites per impurity orbital).\label{fig:bathfit}}
\end{figure}

In quantum impurity models, \(H_{\rm bath}\) and \(H_{\rm hyb}\) are determined by fitting \(\varGamma_{\rm imp}(z)\) to \(\varGamma_{\rm loc}(z) = z - \mathcal{G}_l^{-1}(z)\). For a diagonal hybridization function matrix (ignoring inter-orbital mixing), we have:
\begin{equation}\label{eq:hyb_imp}
  \varGamma_{{\rm imp},l}(z) = \sum_{k} \frac{|V_{lk}|^{2}}{z - \epsilon_{lk}}.
\end{equation}
Fig. \ref{fig:bathfit} presents a fitting example, demonstrating that fitting quality improves significantly with increasing \(n_{\rm b}\). The inset shows that fitting error decreases exponentially with \(n_{\rm b}\); when \(n_{\rm b} \geq 7\), errors for both wide-band (WB) and narrow-band (NB) cases drop below \(10^{-4}\), validating the method's effectiveness.

\subsection{Natural orbital occupancy constraint\label{subsec:NOOC}}

Once a specific impurity model is obtained, the remaining task is to solve it efficiently and accurately. In practice, we find that applying the natural orbital occupancy constraint (NOOC) renders the NORG method for solving impurity models highly efficient and robust~\cite{he2014prb,wang2025abinitio}. Therefore, the computational workflow described below is based on the NOOC-optimized NORG method, abbreviated as NOOC-NORG.

The basic unit of NOOC can be written as \({m}^{l\pm}\), where \(m\) represents the number of spin-orbitals in a group. Because each spin-orbital can host at most one electron, the total electron number in this group is bounded above by \(m\). The notation \(m^*\) denotes an active orbital group, where no occupancy constraints are imposed and the electron number can vary freely from 0 to \(m\). The notation \(l+\) (or \(l-\)) indicates that in the hole constraint group (or electron constraint group), up to \(l\) holes (or electrons) are allowed relative to the fully occupied (or empty) state. Specifically:
\begin{itemize}
  \item For the hole constraint group (\(l+\)), the fully occupied state means all spin-orbitals are occupied by electrons, and allowing up to \(l\) holes implies that the number of electrons is at least \(m - l\);
  \item For the electron constraint group (\(l-\)), the empty state means all spin-orbitals are unoccupied, and allowing up to \(l\) electrons implies that the number of electrons does not exceed \(l\).
\end{itemize}
Once NOOC is determined, we can construct the corresponding Hilbert subspace for the impurity model based on these constraints, and subsequent calculations (such as ground state solving or Green's function computation) will also be performed within this effective Hilbert subspace. 
The NOOC-NORG algorithm is therefore variational: tighter constraints reduce the Hilbert-space dimension, while looser constraints improve accuracy. Choosing an appropriate constraint rule is thus central to solving a given impurity model efficiently.
We devised a series of constraint rules to exploit the sparse impurity-bath interactions and further accelerate NORG by partitioning the Hilbert space more finely: the foundational direct-product constraint, and the progressively more sophisticated joint constraint and particle-hole excitation state (PHES) constraint.


\subsubsection{Direct product constraint rule}

\begin{figure}[htbp!]
  \centering
  \includegraphics[width=10cm]{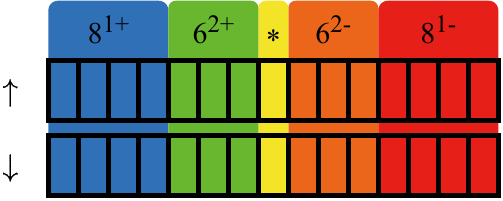}
  \caption{Illustration of the direct-product constraint rule}
  \label{fig:3-4.1}
\end{figure}

We consider the direct product constraint rule using an example of a system where a single impurity orbital couples with 15 bath orbitals. The single impurity corresponds to one spatial orbital (2 spin-orbitals), and the bath corresponds to 15 spatial orbitals (30 spin-orbitals). Thus, the system comprises 32 spin-orbitals in total. For clarity in the following discussion, the term "orbital" will refer to a spin-orbital.

For this illustrative direct-product constraint, the 30 bath spin-orbitals are divided into five groups according to their natural occupations: nearly full, moderately full, active, moderately empty, and nearly empty. The NOOC is applied as follows:
\begin{itemize}
  \item First group (8 orbitals): Relative to full occupancy (all orbitals occupied), up to 1 hole is allowed, denoted as \(8^{1+}\).
  \item Second group (6 orbitals): Relative to full occupancy, up to 2 holes are allowed, denoted as \(6^{2+}\).
  \item Third group (2 orbitals): No occupancy restrictions are applied, denoted as \(2^{*}\). This group's orbital count matches the impurity orbitals. This active group is designed to hybridize effectively with the static impurity orbitals, as it remains unconstrained during the bath transformation.
  \item Fourth group (6 orbitals): Relative to full emptiness (all orbitals unoccupied), up to 2 electrons are allowed, denoted as \(6^{2-}\).
  \item Fifth group (8 orbitals): Relative to full emptiness, up to 1 electron is allowed, denoted as \(8^{1-}\).
\end{itemize}
The core of the direct product constraint rule lies in directly combining each NOOC basic unit via a direct product to construct the total Hilbert subspace of the system. Specifically, the system's Hilbert space is formed by the direct product of the subspaces of the five bath orbital groups and the impurity orbital subspace. The impurity orbital constraint is typically denoted as \(2^{*}\); since the impurity orbitals do not participate in unitary transformation, they are often omitted in notation. Thus, the constraint applied to the bath portion in this example is \(8^{1+}6^{2+}2^{*}6^{2-}8^{1-}\). The five-group split is not imposed by the direct product construction itself; it is a practical grouping chosen for this occupation profile.

\subsubsection{Joint constraint rule}

\begin{figure}[htbp!]
  \centering
  \includegraphics[width=10cm]{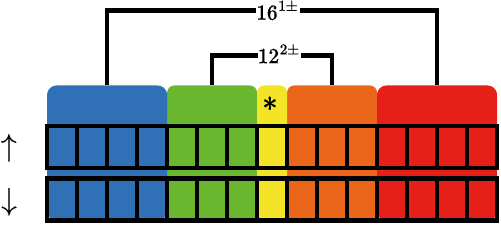}
  \caption{Illustration of the joint constraint rule}
  \label{fig:3-4.2}
\end{figure}

To uniformly handle electron and hole distributions in the impurity model, we designed the joint constraint rule. While the direct product rule is intuitive, it treats electron-like and hole-like excitations in completely separate groups, which can be rigid and computationally suboptimal. The joint constraint rule overcomes this by blurring the distinction between them and introducing the concept of "change quantity." Specifically, the change quantity \(n\) is defined as the number of electrons appearing in regions preset to be fully empty (\(n_e\)) plus the number of holes appearing in regions preset to be fully occupied (\(n_h\)), i.e., \(n = n_e + n_h\).

Under the joint constraint rule, the 30 bath orbitals are divided into three groups:
\begin{itemize}
  \item First group: Contains 2 orbitals, strongly coupled to the impurity orbitals, with no occupancy restrictions, denoted as \(2^{*}\). Since these orbitals are strongly coupled to the impurity, their occupancy distribution significantly affects the system's properties. Therefore, no restrictions are imposed, preserving their full degrees of freedom.
  \item Second group: Contains 12 orbitals, with the first 6 preset to be fully occupied and the latter 6 preset to be fully empty, allowing up to 2 change quantities relative to the preset state (i.e., the sum of holes in the first 6 orbitals and electrons in the latter 6 orbitals does not exceed 2), denoted as \(12^{2\pm}\).
  \item Third group: Contains 16 orbitals, with the first 8 preset to be fully occupied and the latter 8 preset to be fully empty, allowing up to 1 change quantity relative to the preset state (i.e., the sum of holes in the first 8 orbitals and electrons in the latter 8 orbitals does not exceed 1), denoted as \(16^{1\pm}\).
\end{itemize}

Here, the notation \(M^{n\pm}\) indicates a group of \(M\) orbitals where up to \(n\) change quantities are allowed, which can be either electrons in preset fully empty regions or holes in preset fully occupied regions, depending on the orbital's preset state. As shown in Figure \ref{fig:3-4.2}, the division of the three orbital groups and their preset states is illustrated. The innovation of the joint constraint rule lies in uniformly describing holes and electrons through change quantities, offering greater flexibility in adapting to the sparse interaction characteristics in impurity models compared to the traditional direct product constraint rule. Thus, the constraint applied to the bath portion is fully denoted as \(2^{*}12^{2\pm}16^{1\pm}\).

\subsubsection{PHES constraint rule}

\begin{figure}[htbp!]
  \centering
  \includegraphics[width=10cm]{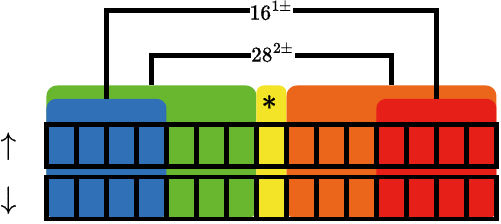}
  \caption{Illustration of the PHES constraint rule}
  \label{fig:3-4.3}
\end{figure}

In studying the joint constraint rule \(2^{*}12^{2\pm}16^{1\pm}\), we found that it could not effectively distinguish configurations with low ground-state weights. To analyze this limitation, we classified all possible configurations by their number of excitations (or "change quantities"), as shown in Table \ref{tab:change_classification}. In the notation \(\begin{bmatrix}* & a & b\end{bmatrix}\) used below, `*' indicates that the number of excitations in the first group is unrestricted, while \(a\) and \(b\) are the numbers of excitations in the second and third groups, respectively; the total number of excitations is \(a+b\).
\noindent\begin{table}[H]
  \centering
  \caption{Classification of configurations by number of excitations under the joint constraint rule}
  \label{tab:change_classification}
  \begin{tabular}{@{}c|c|c|c|c@{}}
    \toprule
    \textbf{No.} & \textbf{Group 1 Excitations} & \textbf{Group 2 Excitations} & \textbf{Group 3 Excitations} & \textbf{Total Excitations} \\
    \midrule
    \#1 & * & 0 & 0 & 0 \\
    \#2 & * & 1 & 0 & 1 \\
    \#3 & * & 2 & 0 & 2 \\
    \#4 & * & 0 & 1 & 1 \\
    \#5 & * & 1 & 1 & 2 \\
    \#6 & * & 2 & 1 & 3 \\
    \bottomrule
  \end{tabular}
\end{table}
Configuration \#6, \(\begin{bmatrix}* & 2 & 1\end{bmatrix}\), has a total of 3 excitations. Although this configuration contributes little to the ground state, the joint constraint rule (\(2^{*}12^{2\pm}16^{1\pm}\)) fails to exclude it. The inclusion of such low-weight, high-excitation configurations reduces computational efficiency.

To address this issue, we introduce the particle-hole excitation state (PHES) constraint rule, which applies a nested set of conditions. This rule uses the same orbital grouping as the joint constraint (an active group, group 2, and group 3), but it imposes two simultaneous conditions on groups 2 and 3:
\begin{enumerate}
    \item A local constraint on the outermost group (group 3, with 16 orbitals), limiting its number of excitations to \(\le 1\).
    \item A global constraint on the combination of group 2 and group 3 (totaling 28 orbitals), limiting their total number of excitations to \(\le 2\).
\end{enumerate}

This nested approach allows the PHES constraint to control configurations more precisely. Consider configuration \#6, \(\begin{bmatrix}* & 2 & 1\end{bmatrix}\). Under the previous joint constraint rule, this configuration is allowed because the number of excitations in group 3 (1) satisfies the \(16^{1\pm}\) limit, and the number in group 2 (2) satisfies the \(12^{2\pm}\) limit. However, the PHES rule rejects it because the total number of excitations in groups 2 and 3 is 3, which violates the global constraint (\(\le 2\)).

Figure \ref{fig:3-4.3} illustrates the orbital grouping and nested structure of the PHES constraint, providing an intuitive visualization of its design. By excluding high-excitation configurations like \#6, the PHES constraint is more effective than the joint constraint at filtering out low-weight configurations, thereby significantly enhancing computational efficiency. Therefore, the complete constraint applied to the bath orbitals is denoted as \(2^{*}(12 + 16^{1\pm})^{2\pm}\).

In summary, \texttt{iNORG} implements a hierarchy of NOOC rules, from the foundational direct product constraint to the more sophisticated joint and PHES constraints. This tiered approach provides the user with the flexibility to select the optimal balance between computational accuracy and efficiency for a given impurity problem. The ability to systematically and physically truncate the vast Hilbert space through these rules is the cornerstone of the NORG method's high performance.

\subsection{Encoding of many-body wave functions\label{subsec:wavefunctioncode}}

To store and rapidly retrieve many-body configurations in quantum impurity solvers, it is crucial to develop efficient encoding methods for Fock states, since the dimension of Hilbert space grows exponentially with the number of orbitals. Here we present the optimized state encoding, which directly supports rapid construction and manipulation of Hamiltonian matrices, essential for iterative methods like the Lanczos and NORG algorithms discussed earlier.

\subsubsection{Combinatorial number encoding}

For any Fock state, it can be viewed as a binary string, where `1' indicates the corresponding orbital is occupied, and `0' indicates empty. For the case of exactly two electrons in six orbitals, a typical scenario encountered in bath orbital encoding within impurity models, the encoding method is demonstrated explicitly. Take \( C(6,2) \) (i.e., \(\binom{6}{2} = 15\), meaning exactly 2 out of 6 orbitals are occupied) as an example, with orbitals numbered from 0 to 5. For instance, \(\ket{0,0,1,1,0,0}\) indicates that both orbitals 2 and 3 (numbering starts from 0) are occupied, corresponding to the binary code ``001100'', which is 12 in decimal. In this way, any configuration satisfying the constraint \(6^{2-}\) (six orbitals, at most two electrons) can be finally converted into an integer code for easy storage in the program.

However, it is inefficient to directly use these 15 decimal bit codes as array indices. As shown in Table~\ref{tab:c62_configurations}, the decimal codes (48, 40, 36, 34, 33, 24, 20, 18, 17, 12, 10, 9, 6, 5, 3) are not consecutive, which would waste memory. A memory-efficient option is to use the combinatorial number system to arrange all states with exactly two occupied orbitals in lexicographical order and map each state to a unique integer rank. This zero-based rank equals the number of states preceding it in lexicographical order. Conversely, given a rank, the Fock state can be reconstructed using the unranking method. The table below lists all 15 configurations and their corresponding information:
\noindent\begin{table}[htbp!]
  \centering
  \caption{Fock states and their encodings for \(C(6,2)\)}
  \label{tab:c62_configurations}
  \begin{tabular}{@{}c|c|c|c|c@{}}
    \toprule
    \textbf{Rank} & \textbf{Occupied orbitals} & \textbf{Fock state} & \textbf{Binary code} & \textbf{Decimal code} \\
    \midrule
    0 & \{0,1\} & $\ket{1,1,0,0,0,0}$ & 110000 & 48 \\
    1 & \{0,2\} & $\ket{1,0,1,0,0,0}$ & 101000 & 40 \\
    2 & \{0,3\} & $\ket{1,0,0,1,0,0}$ & 100100 & 36 \\
    3 & \{0,4\} & $\ket{1,0,0,0,1,0}$ & 100010 & 34 \\
    4 & \{0,5\} & $\ket{1,0,0,0,0,1}$ & 100001 & 33 \\
    5 & \{1,2\} & $\ket{0,1,1,0,0,0}$ & 011000 & 24 \\
    6 & \{1,3\} & $\ket{0,1,0,1,0,0}$ & 010100 & 20 \\
    7 & \{1,4\} & $\ket{0,1,0,0,1,0}$ & 010010 & 18 \\
    8 & \{1,5\} & $\ket{0,1,0,0,0,1}$ & 010001 & 17 \\
    9 & \{2,3\} & $\ket{0,0,1,1,0,0}$ & 001100 & 12 \\
    10 & \{2,4\} & $\ket{0,0,1,0,1,0}$ & 001010 & 10 \\
    11 & \{2,5\} & $\ket{0,0,1,0,0,1}$ & 001001 & 9 \\
    12 & \{3,4\} & $\ket{0,0,0,1,1,0}$ & 000110 & 6 \\
    13 & \{3,5\} & $\ket{0,0,0,1,0,1}$ & 000101 & 5 \\
    14 & \{4,5\} & $\ket{0,0,0,0,1,1}$ & 000011 & 3 \\
    \bottomrule
  \end{tabular}
\end{table}

In Table \ref{tab:c62_configurations}, configurations are sorted in lexicographical order of occupied orbital numbers (from smallest to largest), with numbers 0 to 14 representing their ranks. For example, the configuration \(\ket{0,0,1,1,0,0}\) (occupying orbitals \{2,3\}) has rank 9. The combinatorial ranking/unranking algorithms provide a compact bijection between configurations and integers, significantly enhancing memory and computational efficiency. Using the combinatorial ranking algorithm (Algorithm \ref{alg:ranking}) and unranking algorithm (Algorithm \ref{alg:unranking}), we can achieve bidirectional conversion between configurations and ranks without storing all configurations, thereby saving storage.

\noindent\begin{minipage}[t]{0.49\textwidth}
  \begin{algorithm}[H]
  \caption{Combinatorial ranking algorithm\label{alg:ranking}}
  \begin{algorithmic}[1]
  \STATE \textbf{Input:} comb[1...k] (sorted), n, k
  \STATE \textbf{Output:} rank 
  \STATE rank $\leftarrow 0$, prev $\leftarrow -1$
  \FOR{$i \gets 1$ \TO $k$}
      \FOR{$j \gets$ prev+1 \TO comb[i]-1}
           \STATE rank $\leftarrow$ rank + $C(n-j,\,k-i)$
      \ENDFOR
      \STATE prev $\leftarrow$ comb[i]
  \ENDFOR
  \STATE \textbf{return} rank
  \vspace{1.00\baselineskip}
  \end{algorithmic}
  \end{algorithm}
\end{minipage}\hfill
\begin{minipage}[t]{0.49\textwidth}
  \begin{algorithm}[H]
  \caption{Combinatorial unranking algorithm\label{alg:unranking}}
  \begin{algorithmic}[1]
  \STATE \textbf{Input:} r, n, k \quad ($0 \le r < C(n,k)$)
  \STATE \textbf{Output:} comb[1...k]
  \STATE $r' \leftarrow r$, $x \leftarrow 0$
  \FOR{$i \gets 1$ \TO $k$}
      \WHILE{$C(n-x,\,k-i) \le r'$}
          \STATE $r' \leftarrow r' - C(n-x,\,k-i)$
          \STATE $x \leftarrow x + 1$
      \ENDWHILE
      \STATE comb[i] $\leftarrow x$, $x \leftarrow x + 1$
  \ENDFOR
  \STATE \textbf{return} comb
  \end{algorithmic}
  \end{algorithm}
\end{minipage}
\\
\\
Algorithm \ref{alg:ranking} systematically computes the lexicographical rank of a given occupation configuration, while Algorithm \ref{alg:unranking} reconstructs the occupation configuration from its rank. This approach avoids explicit enumeration, making the storage of configurations highly compact.

\subsubsection{Mixed-radix numeral system}

Building on the combinatorial encoding scheme, the mixed-radix numeral system, a non-standard positional numeral system where each digit can have a different base, further generalizes state representation to efficiently handle multiple orbital groups. The idea is analogous to writing time as days:hours:minutes:seconds, where the second field has base 60, the minute field has base 60, and the hour field has base 24. In the NOOC algorithm, each orbital group plays the role of one digit, and the base of that digit is the number of allowed configurations in that group.

Consider the NOOC constraint \(6^{2+}6^{2-}\), where \(6^{2+}\) means there are at most two holes, while \(6^{2-}\) means there are at most two electrons occupying the six orbitals. The system is divided into two orbital groups, and each group contains 6 orbitals, totaling 12 orbitals. Specifically:
\begin{itemize}
    \item First group: 6 orbitals with \(6^{2+}\) constraint, i.e., relative to the fully occupied state (6 electrons), up to 2 holes are allowed. Possible configurations include:
    \begin{itemize}
        \item No holes: \( C(6,0) = 1 \) (6 electrons);
        \item 1 hole: \( C(6,1) = 6 \) (5 electrons, e.g., \( |1,1,1,0,1,1\rangle \));
        \item 2 holes: \( C(6,2) = 15 \) (4 electrons).
    \end{itemize}
    \item Second group: 6 orbitals with \(6^{2-}\) constraint, i.e., relative to the empty state (0 electrons), up to 2 electrons are allowed. Possible configurations include:
    \begin{itemize}
        \item No electrons: \( C(6,0) = 1 \) (0 electrons);
        \item 1 electron: \( C(6,1) = 6 \) (1 electron);
        \item 2 electrons: \( C(6,2) = 15 \) (2 electrons, e.g., \( |0,0,1,0,0,1\rangle \)).
    \end{itemize}
\end{itemize}

Based on their occupation numbers and indices, each orbital group's configurations can be independently encoded. Let \(A\) and \(B\) be the ranks of the first and second groups, respectively, and let \(R_B\) be the number of allowed configurations in the second group. Then the combined Fock-state code is
\[
n = A \times R_B + B,
\]
with \(0 \le B < R_B\). Thus, \( n \) is a unique integer representing the position of the combined Fock state of the first and second groups in lexicographical order. For the full \(6^{2-}\) constraint, the second group allows zero, one, or two electrons, so its base is \(R_B = C(6,0)+C(6,1)+C(6,2) = 1+6+15 = 22\). For example, if \( A = 3 \) and \( B = 14 \), then \( n = 3 \times 22 + 14 = 80 \). (If the second group were instead restricted to the fixed sector with exactly two electrons, the corresponding base would be \(R_B = C(6,2) = 15\).)

This encoding method maps complex Fock states under NOOC constraints to integers, facilitating storage and retrieval, especially for computational tasks that require traversing all possible configurations.

\subsubsection{Many-body wave function search}

By exploiting physical symmetries, such as U(1) symmetry corresponding to particle-number conservation, the encoding techniques described above enable efficient construction and traversal of the many-body Hilbert space.

For a quantum impurity system, it typically possesses particle number conservation symmetry, i.e., the particle number operator \( \hat{N} \) commutes with the Hamiltonian \( \hat{H} \): \( [\hat{N}, \hat{H}] = 0 \). Therefore, the Hilbert space can be decomposed into distinct subspaces \( \mathcal{H}_N \) with different particle numbers. In practical calculations, we search for many-body wave functions within a specific subspace \( \mathcal{H}_N \).

Take a system with 8 spatial orbitals (16 orbitals) as an example, considering the NOOC constraint \(6^{2+}4^{*}6^{2-}\), with orbitals allocated as follows:
\begin{itemize}
    \item First group: 6 orbitals with \(6^{2+}\) constraint (up to 2 holes);
    \item Second group: 4 orbitals with \(4^{*}\) constraint (no occupancy constraints), composed of the 2 impurity orbitals and 2 bath orbitals for this example;
    \item Third group: 6 orbitals with \(6^{2-}\) constraint (up to 2 electrons).
\end{itemize}
Then the total number of orbitals is \( 6 + 4 + 6 = 16 \).

In the half-filling (particle number \( N = 8 \)) subspace \( \mathcal{H}_8 \), we use NOOC constraints and particle number conservation to determine the distribution of many-body wave functions:
\begin{itemize}
    \item First group \(6^{2+}\): Electron number \( N_1 \) can be 4, 5, 6 (corresponding to 2, 1, 0 holes), with configuration numbers \( C(6,4) \), \( C(6,5) \), \( C(6,6) \), i.e., 15, 6, 1.
    \item Second group \(4^{*}\): Electron number \( N_2 = N - N_1 - N_3 = 8 - N_1 - N_3 \), ranging from 0 to 4, with configuration number \( C(4, N_2) \).
    \item Third group \(6^{2-}\): Electron number \( N_3 \) can be 0, 1, 2 (corresponding to 0, 1, 2 electrons), with configuration numbers \( C(6,0) \), \( C(6,1) \), \( C(6,2) \), i.e., 1, 6, 15.
\end{itemize}
By traversing all possible combinations of \( N_1 \) and \( N_3 \), we obtain \( N_2 \), and then check if it is between 0 and 4. For example, if \( N_1 = 5 \) (1 hole, configuration number \( C(6,5) = 6 \)) and \( N_3 = 1 \) (1 electron, configuration number \( C(6,1) = 6 \)), then \( N_2 = 8 - 5 - 1 = 2 \), with configuration number \( C(4,2) = 6 \). For this specific distribution of electrons, the total number of configurations is given by the product of those from each group, yielding \(6 \times 6 \times 6 = 216\). 

In summary, by combining combinatorial ranking and mixed radix encoding schemes, we significantly reduce the computational resources required for encoding and traversing the many-body Hilbert space, thereby enhancing the overall efficiency of the solver.

\subsection{Rapid generation of many-body operators\label{subsec:operator}}

For computationally feasible treatments of quantum impurity problems, the rapid and efficient generation of many-body operators is crucial,  especially in systems characterized by large and sparse Hamiltonian matrices. In this section, we present the key techniques employed in \texttt{iNORG} to address these computational challenges.

\subsubsection{Hamiltonian construction}

In practical calculations, the dimension of the Hilbert space can be very large (e.g., up to 160 million dimensions), making it impractical to directly construct and store the complete Hamiltonian matrix \( H_{mn} = \langle m | \hat{H} | n \rangle \) (where \(\{|n\rangle\}\) is the Fock state basis). To address the storage limitations and matrix sparsity, we adopt the following efficient method:

We first apply \(\hat{H}\) to a reference Fock state \(\ket{n}\) and enumerate all states \(\ket{m}\) for which \(\langle m | \hat{H} | n \rangle \neq 0\). For a Hamiltonian of the form $\hat{H}=\sum_{ij}t_{ij}\hat{c}^\dagger_i \hat{c}_j+\sum_{ijkl}U_{ijkl}\hat{c}^\dagger_i \hat{c}^\dagger_j \hat{c}_k \hat{c}_l$, the action of these operators on a Fock state reveals all accessible transitions. This approach efficiently pinpoints non-zero elements and their corresponding row and column indices.

We store only the non-zero elements in compressed sparse row (CSR) format \cite{intel2024sparse}, whose three-array structure (row pointers, column indices, values)\footnotemark enables fast sparse matrix-vector products required by iterative algorithms such as Lanczos.

\footnotetext{
For clarity, consider an example:
\[
\mathbf{A} = \begin{bmatrix}
1 & 0 & 0 \\
0 & 2 & 0 \\
3 & 0 & 4
\end{bmatrix}
\]
Its CSR format consists of three arrays: row pointers (row\_ptr): [0, 1, 2, 4], column indices (col\_ind): [0, 1, 0, 2], values (values): [1, 2, 3, 4].
}

\subsubsection{NORG-Table data structure}

To avoid regenerating the CSR matrix from scratch in every Hamiltonian update, we introduce the NORG-Table. For every non-zero element \(\langle m|\hat{H}|n\rangle\), the table stores one row \((m,n,t_{\mathrm{tag}})\), where \(t_{\mathrm{tag}}\) is a compact integer that identifies the single-particle operator linking \(|n\rangle\) and \(|m\rangle\).

Let \(N_{\mathrm{sp}}\) denote the number of single-particle spin-orbitals. We label operators as
\begin{equation}
t_{\mathrm{tag}} = 
\begin{cases}
(i-1)N_{\mathrm{sp}} + (j-1), & \hat{c}^{\dagger}_{i}\hat{c}_{j} \\[4pt]
N_{\mathrm{sp}}^{2} + \{[(i-1)N_{\mathrm{sp}} + (j-1)]N_{\mathrm{sp}} + (k-1)\}N_{\mathrm{sp}} + (l-1), &
\hat{c}^{\dagger}_{i}\hat{c}^{\dagger}_{j}\hat{c}_{k}\hat{c}_{l}
\end{cases}
\end{equation}
ensuring a unique range \(0\le t_{\mathrm{tag}}<N_{\mathrm{sp}}^{2}+N_{\mathrm{sp}}^{4}\). The table must be rebuilt when the NOOC subspace or orbital basis changes. Within that fixed structure, updated Hamiltonian coefficients can reuse the same row and tag information, so the corresponding matrix elements can be reconstructed without another traversal of the Hilbert space.

During NORG iterations, the Table provides direct access to non-zero matrix elements, avoiding repeated traversals of the Hilbert space whenever only the numerical values of the single-particle or interaction parameters are updated. The recorded information can be directly used to construct sparse matrices in CSR format, the standard format for storing the sparse matrix of \(\hat{H}\), which enables rapid reconstruction of matrix elements in DMFT iterations, thereby significantly enhancing computational efficiency.

\subsection{Ground-state solution based on the Lanczos algorithm\label{subsec:lanc_gs}}

At every NORG sweep, we must determine the ground-state vector of the current effective Hilbert subspace. To achieve this aim efficiently, we adopt the Lanczos method. The resulting pseudocode is given in Algorithm \ref{alg:lanczos}, where \(\gets\) denotes assignment.

\begin{algorithm}[H]
  \caption{Lanczos ground-state solver \label{alg:lanczos}}
  \begin{algorithmic}[1]
  \REQUIRE Hamiltonian \( H \), initial normalized state \( \varphi_0 \), maximum Krylov dimension \( m \)
  \STATE Set \( \mathbf{w}_0 \gets \varphi_0 \)
  \STATE Compute \( \mathbf{w}_1 \gets H\, \mathbf{w}_0 \)

  \FOR{$i = 1$ \TO $m$}
      \STATE Compute \( \alpha_{i-1} \gets \langle \mathbf{w}_0,\, \mathbf{w}_1 \rangle \)
      \IF{\{Lanczos convergence criterion [see \eqref{eq3:lanczosResidual}]\}}
          \STATE \textbf{break}
      \ENDIF
      \STATE Update \( \mathbf{w}_1 \gets \mathbf{w}_1 - \alpha_{i-1}\, \mathbf{w}_0 \)
      \STATE Compute \( \beta_i \gets \| \mathbf{w}_1 \| \)
      \STATE Normalize \( \mathbf{w}_1 \gets \dfrac{\mathbf{w}_1}{\beta_i} \)
      \STATE \textbf{Swap} \( \mathbf{w}_0 \) and \( \mathbf{w}_1 \) 
      \STATE Update \( \mathbf{w}_1 \gets H\, \mathbf{w}_0 - \beta_i\, \mathbf{w}_1 \)
  \ENDFOR
  
  \RETURN The sequences \( \{\alpha_0, \alpha_1, \dots, \alpha_{m-1}\} \) and \( \{\beta_1, \beta_2, \dots, \beta_{m-1}\} \)
  \end{algorithmic}
  \end{algorithm}

In the above iterative process, to further determine whether the Lanczos iteration has converged, the concept of "residual" is often introduced. Simply put, if we diagonalize the tridiagonal matrix obtained after \( i \) iterations (with diagonal elements \(\{\alpha_0, \alpha_1, \dots, \alpha_{i-1}\}\), and off-diagonal elements \(\{\beta_1, \beta_2, \dots, \beta_{i-1}\}\)), we can obtain a series of approximate eigenvalues and eigenvectors (commonly referred to as Ritz values and Ritz vectors). Among them, the Ritz vector corresponding to the lowest energy (ground-state) approximation can be denoted as \(\boldsymbol{y} = (y_1, y_2, \dots, y_i)^T\), and the residual norm of this state under the original Hamiltonian \cite{golub2013book} reads
\begin{equation}
\|H\,\mathbf{v} - \theta\,\mathbf{v}\|_2 = \beta_i \, \bigl|\,y_i\bigr|,
\label{eq3:lanczosResidual}
\end{equation}
where \(\theta\) is the Ritz value corresponding to the approximate state, \(\mathbf{v}\) is the representation of the approximate state in the original space, and \(\beta_i\) is the last off-diagonal element obtained in this iteration. Generally, if \(\beta_i \, |\,y_i|\) is less than a certain tolerance (e.g., \(\epsilon\)), namely 
\begin{equation}
  \beta_i \, \bigl|\,y_i\bigr| \le \epsilon,
  \label{eq3:ResidualCriterion}
\end{equation}
it can be considered that the Krylov subspace is sufficient to approximate the ground state.

In the criterion \eqref{eq3:ResidualCriterion}, we rely on the product of the last off-diagonal element \(\beta_i\) and the magnitude of the last component of the corresponding Ritz vector \(|y_i|\) as the convergence criterion. When the value is sufficiently small, it indicates that the current Krylov subspace contains sufficient information to approximate the ground state, and the convergence is considered to be achieved. This criterion provides reliable convergence control without materially increasing cost.

It is emphasized that numerical errors and stability are crucial issues in this algorithm; if not controlled, numerical errors can accumulate. The residual criterion monitors convergence of a Ritz pair, but it does not by itself prevent finite-precision Krylov vectors from gradually losing orthogonality. Typically, methods such as Kahan summation (to reduce floating-point error accumulation) and optional partial re-orthogonalisation (to restore orthogonality when needed) are used to control errors \cite{wu1999jcp, sundar2000cma, wu2000siam, kokiopoulou2004anm}. A small spectral shift can place zero energy near the ground-state level, accelerating convergence. If the system has many degenerate eigenstates, block Lanczos algorithms (solving multiple eigenstates simultaneously) or increasing the Krylov dimension may be more appropriate.

\subsection{Continued-fraction evaluation of diagonal Green's functions\label{subsec:continuefraction}}

For quantum many-body problems, the Green's function encodes the spectral weight, elementary excitations, and other dynamical observables. Once the ground state is available, dynamical properties can be obtained from the single-particle Green's function, which we will evaluate efficiently with a Lanczos-based continued-fraction scheme. In this subsection, we will introduce how to use the continued-fraction inversion algorithm to solve the zero-temperature single-particle (retarded) Green's function \cite{gagliano1987prl}. We first review the basic formula of the Green's function, then detail how to construct the tridiagonal matrix during the Lanczos iteration process, and finally use the continued-fraction expansion inversion method to solve the Green's function numerically.

The formula for the zero-temperature retarded Green's function is expressed as follows,
\begin{align}
  G_{ij\sigma}^{\mathrm{R}}(\omega + i\eta) &= G_{ij\sigma}^{>}(\omega + i\eta) + G_{ij\sigma}^{<}(\omega + i\eta) \notag\\[1mm]
  &= \langle\psi_0| c_{i\sigma} \frac{1}{\omega + i\eta + E_0 - H} c_{j\sigma}^{\dagger} |\psi_0\rangle + \langle\psi_0| c_{j\sigma}^{\dagger} \frac{1}{\omega + i\eta - E_0 + H} c_{i\sigma} |\psi_0\rangle
\end{align}
where $\eta > 0$ is an infinitesimal broadening factor. In practice, direct inversion or diagonalization is infeasible, due to the typically enormous dimensionality of the Hamiltonian \(H\). Therefore, Lanczos iteration constructs a tridiagonal representation \(T\) of \(H\) in the Krylov space, thereby transforming the inversion problem into a finite-dimensional subspace. This is exactly the Lanczos continued-fraction method for computing the Green's function. Taking
\begin{equation}
  G^{<}(z) = \langle\psi_0| c^{\dagger} \frac{1}{z + H - E_0} c |\psi_0\rangle
\end{equation}
as an example, we detail its computational implementation below.

We first define the initial state as \(|f_0\rangle = c|\psi_0\rangle\), and introduce two sets of Krylov complete bases \(\{\ket{n_i}\}\) and \(\{\ket{n_j}\}\), then the Green's function \(G^{<}(z)\) can be written as
\begin{align}
  G^{<}(z) &= \sum_{i,j} \braket{f_0}{n_i}\,\bra{n_i}\,\frac{1}{z+H-E_0}\,\ket{n_j}\,\braket{n_j}{f_0} \notag\\[1mm]
           &= \sum_{i,j} \delta_{i,0}\,\bra{n_i} \,\frac{1}{z+H-E_0}\,\ket{n_j} \,\delta_{j,0} \notag\\[1mm]
           &= \left[\frac{1}{z+T-E_0}\right]_{00},
\end{align}
where \(\delta_{i,0}\) is 1 when \(i=0\) and 0 otherwise, indicating that \(|f_0\rangle\) coincides with the first basis vector in the Krylov basis. Here, the subscript "00" denotes the first element of the matrix \(\left[(z+T-E_0)^{-1}\right]\). During the Lanczos iteration process, we have the following formulas:
\[
\begin{aligned}
\ket{n_0} &= \frac{c\ket{\psi_0}}{\|c\ket{\psi_0}\|}, \quad
\bra{n_0} = \frac{\bra{\psi_0}c^\dagger}{\|c\ket{\psi_0}\|}, \\
\beta_0 &= 0, \\
\alpha_i &= \bra{n_i} H \ket{n_i}, \\
\beta_{i+1} \ket{n_{i+1}} &= \bigl(H-\alpha_i\bigr)\ket{n_i}-\beta_i\ket{n_{i-1}}.
\end{aligned}
\]
Specifically, the diagonal elements of \(T\) are \(\{\alpha_i\}\), and the upper (lower) off-diagonal elements are \(\{\beta_{i+1}\}\). Using Algorithm \ref{alg:lanczos}, we can project the operator \(H\) into the corresponding Krylov space to obtain the tridiagonal matrix \(T\) with
\begin{equation}
  T =
  \begin{pmatrix}
    \alpha_0 & \beta_1 & 0       & \cdots & 0\\[1mm]
    \beta_1  & \alpha_1 & \beta_2 & \cdots & 0\\[1mm]
    0        & \beta_2  & \alpha_2 & \cdots & 0\\[1mm]
    \vdots   & \vdots   & \vdots  & \ddots & \beta_{N-1}\\[1mm]
    0        & 0        & 0       & \beta_{N-1} & \alpha_{N-1}
  \end{pmatrix}.
\end{equation}
Because T is tridiagonal, this matrix element is expressible as an infinite continued fraction.
Using the continued-fraction method \cite{haydock1975jpc}, we can compute \(\left[(z+T-E_0)^{-1}\right]_{00}\), i.e., the value of the first element in the upper-left corner of the inverse matrix, expressed in continued-fraction form as:
\begin{align}
  G^{<}(z) & = \bigl[(z + T - E_0)^{-1}\bigr]_{00} \notag\\
  &=\; 
  \cfrac{1}{z - E_0 + \alpha_0 \;-\; 
  \cfrac{\beta_1^2}{z - E_0 + \alpha_1 \;-\; 
  \cfrac{\beta_2^2}{z - E_0 + \alpha_2 \;-\; \dots}}},
\end{align}
Similarly, we have:
\begin{align}
  G^{>}(z) &= \bigl[(z - T + E_0)^{-1}\bigr]_{00} \notag\\
  &=\; 
  \cfrac{1}{z + E_0 - \alpha_0 \;-\; 
  \cfrac{\beta_1^2}{z + E_0 - \alpha_1 \;-\; 
  \cfrac{\beta_2^2}{z + E_0 - \alpha_2 \;-\; \dots}}},
\end{align}
By appropriately truncating at a finite order (e.g., continue until $|\beta_N| < \epsilon$ or up to a prescribed Krylov dimension $N$), we can obtain a numerical approximation to \(G^{<}(z)\).

Thus, by projecting $H$ onto a Lanczos basis, and evaluating the first diagonal element of $(z+T-E_0)^{-1}$ through its continued fraction, it yields the zero-temperature single-particle Green's function with $O(N)$ memory and $O(N^2)$ time.

\subsection{Krylov-subspace evaluation of matrix-valued Green's functions\label{subsec:krylovsubspace}}

While the continued-fraction Lanczos technique is well suited for a single vector $\langle v|(z-H)^{-1}|v\rangle$, it cannot directly deliver generic matrix elements $\langle u|(z-H)^{-1}|v\rangle$. Therefore, we need algorithms that build a Krylov space capable of treating multiple source and sink states simultaneously. Two practical remedies are widely used: one is the "correction vector" algorithm, and the other is the Green's function algorithm based on the Krylov subspace.

The correction-vector method rewrites the resolvent $(\omega + i\eta + E_0 - H)^{-1}$ as a linear system, which is solved independently for every frequency point $\omega$. Specifically, a correction vector \(|\chi(\omega)\rangle\) is introduced, satisfying
\begin{equation}
  (\omega + i\eta + E_0 - H)\,|\chi(\omega)\rangle \;=\; A\,|\psi_0\rangle,
\end{equation}
where \(A\) is the corresponding annihilation/creation operator, \(|\psi_0\rangle\) is the system's ground state, and \(\eta\) represents the energy-broadening parameter. After numerically solving for \(|\chi(\omega)\rangle\), the Green's function can be obtained from the simple inner product \(\langle \psi_0| A^\dagger | \chi(\omega) \rangle\). This method allows direct computation in the frequency domain without Fourier transformation, and it is often applied to study dynamical properties of strongly correlated electron systems employing many-body numerical methods, such as the DMRG method\cite{kuhner1999prb, peters2011prb, dargel2012prb}. However, the cost is dominated by an independent linear solve for every source operator and every one of the $N_\omega$ sampled frequencies, with each solve acting on the full Hilbert space.

Since the correction-vector method requires multiple solutions in practice, it leads to a large computational load and is generally not ideal for large-scale strongly correlated models. In a matrix-valued Green's function \(G_{\alpha\beta}(\omega)=\langle \phi'_{\alpha}|(z-H)^{-1}|\phi_{\beta}\rangle\), \(|\phi_{\beta}\rangle\) is the source vector associated with column \(\beta\), and \(\langle\phi'_{\alpha}|\) is the sink vector associated with row \(\alpha\). The Krylov-subspace approach builds one projected (Krylov) space for each source vector and reuses that space for all sink vectors and all sampled frequencies, reducing redundant computations and improving scalability. We define
$$
z \equiv E_0 + \omega + i\eta, \qquad \eta > 0,
$$
and consider the particle part of the zero-temperature Green's function

\begin{align}
G^{>}(\omega)
  &= \langle \phi' | (z-H)^{-1} | \phi \rangle \nonumber\\
  &\approx \langle \phi' | V (z-T)^{-1} V^{\dagger} | \phi \rangle \nonumber\\
  &= \langle \phi' | V S (z-D)^{-1} S^{\dagger} V^{\dagger} | \phi \rangle \nonumber\\
  &= \langle n | (z-D)^{-1} | m \rangle,
\end{align}

where
$$
|m\rangle = S^{\dagger} V^{\dagger}|\phi\rangle, \qquad
\langle n| = \langle \phi'|V S.
$$
Here \(V\) contains the orthonormal Krylov basis vectors as its columns and \(T=V^{\dagger}HV\) is the projected tridiagonal Hamiltonian. The unitary matrix \(S\) diagonalizes \(T\), \(S^{\dagger}TS=D\), so that \((z-T)^{-1}=S(z-D)^{-1}S^{\dagger}\). After the projected source and sink vectors are rotated by \(S^{\dagger}\) and \(S\), respectively, the projected resolvent is represented in the eigenbasis of \(T\) by the diagonal matrix
\[
M(z) \equiv (z-D)^{-1}.
\]
The overall computational process can be divided into three numbered steps:

\begin{enumerate}
\item Lanczos projection - build \(V\), compute \(T=V^{\dagger}HV\), and project the source and sink vectors onto the Krylov space, \(V^{\dagger}|\phi\rangle\) and \(\langle\phi'|V\).
\item Diagonalisation - find $S^{\dagger}TS=D$ and rotate the projected vectors into the eigenbasis, \(|m\rangle=S^{\dagger}V^{\dagger}|\phi\rangle\) and \(\langle n|=\langle\phi'|VS\).
\item Assembly - for each complex $z$, with the diagonal $M(z)=(z-D)^{-1}$, form the matrix elements $G_{\alpha\beta}(\omega)=\langle n_\alpha|M(z)|m_\beta\rangle=\sum_{k}(n_\alpha)_k (m_\beta)_k/(z-D_k)$.
\end{enumerate}

Because storing all full Hilbert-space source vectors is prohibitive, we instead retain the projected quantities $|m\rangle$ and $\langle n|$, whose dimension equals the Krylov size and is much smaller than the full Hilbert-space dimension. For an \(N_{\mathrm{orb}} \times N_{\mathrm{orb}}\) Green's function matrix, there are \(N_{\mathrm{orb}}\) different single-particle source states; each is treated in a separate Lanczos run and projected onto the corresponding Krylov space. Projecting onto the Krylov basis and rotating into the eigenbasis of \(T\), we obtain
\begin{align*}
  |m\rangle & =S^{\dagger}V^{\dagger}|\phi\rangle,\\
  \langle n| & =\langle \phi'|VS.
\end{align*}
It is important to note that one Lanczos run uses the same \(V\) and \(S\), so it can simultaneously generate one \(|m\rangle\) and \(N_{\mathrm{orb}}\) projected sink vectors \(\langle n|\), which together yield a single column of the Green's function matrix; the full matrix therefore requires \(N_{\mathrm{orb}}\) such runs. When computing the Green's function matrix later, we only need the inner product \(\left\langle n\right| M(z) \left|m\right\rangle\) to obtain the Green's function value.

Across the \(N_{\mathrm{orb}}\) source-resolved Lanczos runs, once the projected quantities \(V\), \(T\) and \(S\) have been produced, assembling all \(N_{\mathrm{orb}}^{2}\) matrix elements for \(N_\omega\) frequency points costs only \(O(N_{\text{K}} N_{\mathrm{orb}}^{2} N_\omega)\) arithmetic, where \(N_{\text{K}}\) is the Krylov dimension. The decisive advantage is that this per-frequency assembly uses only Krylov-dimension operations, independent of the full Hilbert-space dimension \(N_{\text{H}}\), whereas the correction-vector method repeats a full Hilbert-space solve at every frequency and scales as \(O(N_{\text{H}} N_{\mathrm{orb}} N_\omega)\). The hole part \(G^{<}\) is obtained in complete analogy. This approach not only reduces the numerical complexity of computing large-scale matrices significantly, but also has important applications in the study of spectral properties of strongly correlated systems.


\section{Installation and usage\label{sec:install}}

This section details the steps required to obtain, build, and run the \texttt{iNORG} software package.

\subsection{Obtaining iNORG\label{subsec:get}}
\texttt{iNORG} is an open-source software package released under the AGPLv3 license. You can obtain the source code by cloning the public repository at \url{https://github.com/QTMEC-RUC/iNORG}:
\begin{verbatim}
    $ git clone https://github.com/QTMEC-RUC/iNORG.git
    $ cd iNORG
\end{verbatim}

\subsection{Prerequisites\label{subsec:prereq}}
Before building \texttt{iNORG}, ensure the following dependencies are installed and configured on your system:
\begin{itemize}
    \item C++ Compiler: A modern compiler supporting the C++17 standard (e.g., GCC 7+, Clang 6+, Intel C++ Compiler 19+).
    \item MPI Library: An MPI implementation such as OpenMPI, MPICH, or Intel MPI. The corresponding MPI compiler wrappers (\texttt{mpicxx} or similar) should be available in your PATH.
    \item Intel Math Kernel Library (MKL): Required for optimized BLAS and LAPACK routines. Ensure the MKL environment variables are set correctly (e.g., via \texttt{mklvars.sh}).
\end{itemize}

\subsection{Building \texttt{iNORG}\label{subsec:build}}
Follow these steps to compile the \texttt{iNORG} executable:
\begin{enumerate}
    \item Create a build directory and navigate into it:
    \begin{verbatim}
        $ mkdir build
        $ cd build
    \end{verbatim}
    \item Configure the build using Make. You might need to specify the path to MKL if it's not found automatically. Replace \texttt{/path/to/mkl} with your MKL installation root.
    \begin{verbatim}
        $ make clean  # Optional, but recommended for clean build
        $ make
    \end{verbatim}
    \textit{Note: Ensure \texttt{mpicxx} points to your desired MPI-enabled C++ compiler.}
    \item Compile the code using Make:
    \begin{verbatim}
        $ make -j N  # Replace N with the number of parallel jobs
    \end{verbatim}
\end{enumerate}
Upon successful compilation, the executable (e.g., \texttt{inorg}) will typically be located in the \texttt{build/src} or \texttt{build/bin} directory.

\subsection{Running \texttt{iNORG}\label{subsec:run}}
To run an \texttt{iNORG} calculation:
\begin{enumerate}
    \item Prepare Input Files: Create the necessary input files in your working directory. This typically includes:
        \begin{itemize}
            \item A main parameter file (e.g., \texttt{PARAMS.norg}) specifying the model, solver settings, convergence criteria, etc.
            \item Files containing the initial hybridization function \(\varGamma(i\omega_n)\) if starting a DMFT cycle or using a predefined bath.
        \end{itemize}
        Refer to the documentation or examples for the specific format and required parameters.
    \item Set Up Environment (if needed): Ensure the MKL runtime libraries are accessible. This might involve sourcing the \texttt{mklvars.sh} script in your run script or environment:
    \begin{verbatim}
        $ source /path/to/mkl/bin/mklvars.sh intel64  # Adjust path and architecture
    \end{verbatim}
    \item Execute: Run the program using \texttt{mpirun} or your system's equivalent MPI launcher. Specify the number of MPI processes (\texttt{num\_procs}) and the path to the \texttt{iNORG} executable.
    \begin{verbatim}
        $ mpirun -n <num_procs> /path/to/iNORG/build/src/inorg > output.log
    \end{verbatim}
    Redirecting output to a log file (\texttt{output.log}) is recommended.
    \item Analyze Output: Monitor the log file for convergence information and check the output files generated by \texttt{iNORG} (e.g., calculated Green's functions, self-energies, natural orbital occupations).
\end{enumerate}

Important notes:
\begin{itemize}
    \item The number of MPI processes should be chosen based on your system resources and problem size
    \item For large systems, it's recommended to first test with a small number of processes
    \item Check the detailed documentation and example files in the doc/ directory for specific usage scenarios
    \item The program supports checkpoint/restart functionality for long calculations
\end{itemize}

For detailed information about input file formats, available parameters, and advanced usage options, please refer to the user manual in the documentation.

\section{Examples\label{sec:exp}}

This section presents two practical examples demonstrating \texttt{iNORG}'s capabilities for solving quantum impurity models. The first example shows a one-shot solution of a multi-orbital impurity problem within DMFT, while the second demonstrates a self-consistent multi-orbital DMFT calculation on a Bethe lattice. Each example includes complete parameter files with detailed explanations and execution instructions.

\subsection{Example 1: One-shot solution of multi-orbital impurity problem in DMFT\label{subsec:exp1}}

This example demonstrates solving a three-orbital Anderson impurity model with Coulomb interactions in the context of DMFT. All energy parameters are given in eV units.

\noindent\textbf{Parameter file (\texttt{PARAMS.norg}):}

\begin{lstlisting}[language=norg,
  basicstyle=\ttfamily\small,
  backgroundcolor=\color{yellow!10},
  commentstyle=\color{olive!10!green},
  keywordstyle=\color{purple}]
# Three-orbital Anderson impurity model for DMFT
# All energy parameters in eV units

# Model configuration
mode            1
Ed              [-11.61347089, -11.62251229, -11.63870539]
deg_idx         [1, 2, 3]

# Interaction parameters  
U               8.0
J               0.8
CoulombF        'Ising'

# Temperature and fitting
beta            200.0
fit_points      [0, 1, 2, 3, 4, 5, 7, 10, 13, ..., 1088, 2164, 3780]
fit_nbaths      [11, 11, 11, 11, 11, 11]

# NOOC configuration
NOOC            phss_v2
restrain        [0, -2, -3, -4, 0, 4, 3, 2]
pred_gs_deg     2
ful_pcl_sch     1
weight_nooc     [1e-05, 1e-05, 1e-05, 1e-05, 1e-05]
weight_freze    [1e-07, 1e-07, 1e-07, 1e-07, 1e-07]
\end{lstlisting}

\noindent\textbf{Parameter descriptions:}
\begin{itemize}
\item \texttt{mode}: Solver mode. \texttt{1} indicates a one-shot solution of the impurity model.

\item \texttt{Ed}: On-site energies of the impurity levels (in eV).

\item \texttt{deg\_idx}: Degeneracies corresponding to each impurity level in \texttt{Ed}.

\item \texttt{U}: Intra-orbital Coulomb repulsion parameter $F_0 = 8.0$ eV.

\item \texttt{J}: Hund's coupling strength $J = 0.8$ eV.

\item \texttt{CoulombF}: Form of Coulomb interaction. The current version supports only the density-density (\texttt{Ising}) form.

\item \texttt{beta}: Inverse temperature $\beta = 1/(k_B T) = 200$ eV$^{-1}$ (corresponding to $T \approx 0.005$ eV).

\item \texttt{fit\_points}: Discrete Matsubara frequency points for hybridization function fitting, selected using the discrete Lehmann representation \cite{kaye2022prb,kaye2022cpc} with $E_{\text{uv}} = 10$, $\beta = 200$, and relative tolerance $10^{-6}$.

\item \texttt{fit\_nbaths}: Number of bath sites for each irreducible representation (six sectors, each with 11 bath sites).

\item \texttt{NOOC}: Type of Natural-Orbital Occupancy Constraint applied (\texttt{phss\_v2} scheme, see Section~\ref{subsec:NOOC}).

\item \texttt{restrain}: NOOC distribution parameters among orbital groups. The eight values correspond to different orbital group constraints as detailed in Section~\ref{subsec:NOOC}.

\item \texttt{pred\_gs\_deg}: Expected ground-state degeneracy (2). If set to 0, degeneracy checking is skipped.

\item \texttt{ful\_pcl\_sch}: Enables (1) or disables (0) particle-number subspace screening.

\item \texttt{weight\_nooc}, \texttt{weight\_freze}: Thresholds for filtering natural-orbital weights in the NOOC procedure.
\end{itemize}

\noindent\textbf{Execution:}
\begin{verbatim}
$ mpirun -n 4 /path/to/iNORG/build/src/inorg > output_example1.log
\end{verbatim}

\subsection{Example 2: Multi-orbital DMFT calculation on Bethe lattice\label{subsec:exp2}}

This example demonstrates a self-consistent DMFT calculation for a two-orbital Bethe lattice model with inter-orbital interactions. Energy parameters are in dimensionless units relative to the hopping amplitude.

\noindent\textbf{Parameter file (\texttt{PARAMS.norg}):}

\begin{lstlisting}[language=norg,
  basicstyle=\ttfamily\small,
  backgroundcolor=\color{yellow!10},
  commentstyle=\color{olive!10!green},
  keywordstyle=\color{purple}]
# Two-orbital Bethe lattice DMFT calculation
# Energy parameters in dimensionless units (relative to hopping)

# Model configuration
mode           0
bethe_mu       0.0
bethe_band     2
bethe_t        [0.5, 0.25]

# Bath fitting
fit_nbaths     [7, 7, 7, 7]

# Interaction parameters
bethe_u        3.0
bethe_uprim    2.7

# NOOC configuration  
pred_gs_deg    2
NOOC           phss_v2
restrain       [0, -2, -3, -4, 0, 4, 3, 2]
distribute     [1, 0, 1, 2, 1, 2, 1, 0]
\end{lstlisting}

\noindent\textbf{Parameter descriptions:}

Parameters identical to Example 1 (\texttt{NOOC}, \texttt{restrain}) are not repeated here.

\begin{itemize}
\item \texttt{mode}: Solver mode. \texttt{0} indicates a full self-consistent DMFT loop.

\item \texttt{bethe\_mu}: Chemical potential for the Bethe lattice (set to 0 for half-filling).

\item \texttt{bethe\_band}: Number of orbitals per site (2 for this two-orbital model).

\item \texttt{bethe\_t}: Hopping amplitudes for each orbital. The bandwidth is $4t$ for each orbital.

\item \texttt{fit\_nbaths}: Number of bath sites for each irreducible representation (four sectors with 7 bath sites each).

\item \texttt{bethe\_u}: Intra-orbital Coulomb interaction strength $U = 3.0$.

\item \texttt{bethe\_uprim}: Inter-orbital Coulomb interaction strength $U' = 2.7$.

\item \texttt{pred\_gs\_deg}: Expected ground-state degeneracy (2).

\item \texttt{distribute}: Template defining orbital positions within NOOC groups. The eight values map specific orbitals to the constraint groups defined by \texttt{restrain}.
\end{itemize}

\noindent\textbf{Execution:}
\begin{verbatim}
$ mpirun -n 8 /path/to/iNORG/build/src/inorg > output_example2.log
\end{verbatim}

\section{Future developments\label{sec:future}}
Future enhancements for \texttt{iNORG} are planned across several key areas to expand its capabilities and improve its performance and usability:
\begin{enumerate}
    \item Advanced features including adaptive NOOC schemes and finite-temperature extensions.
    \item Support for two-particle responses and cluster DMFT.
    \item Python interfaces for integration with existing DFT+DMFT ecosystems.
\end{enumerate}
These developments will solidify \texttt{iNORG} as a powerful and versatile tool for cutting-edge research in strongly correlated electron systems.

\vspace{10pt}

\noindent \textbf{Declaration of competing interest}

\vspace{5pt}
The authors declare that they have no known competing financial interests or personal relationships that could have appeared to influence the work reported in this paper.\\

\noindent \textbf{Code availability}

\vspace{5pt}
\texttt{iNORG} is free and open-source software, released under the GNU Affero General Public License v3.0 (AGPL-3.0). The complete source code, build instructions, and worked examples reproducing the results reported in this paper are publicly available in the developer repository at \url{https://github.com/QTMEC-RUC/iNORG}.\\

\noindent \textbf{Acknowledgement}

\vspace{5pt}
This work was supported by the National Key R\&D Program of China (Grants No. 2024YFA1408602 and No. 2024YFA1408601) and the National Natural Science Foundation of China (Grant No. 12434009). Z.-Y.L. was also supported by the Innovation Program for Quantum Science and Technology (Grant No. 2021ZD0302402). Computational resources were provided by the Physical Laboratory of High Performance Computing in Renmin University of China.

\bibliographystyle{unsrt}
\bibliography{inorg}

\end{document}